\documentclass[12pt,notitlepage,a4paper]{article}
\pdfoutput=1
\usepackage{color,graphicx}
\usepackage{amssymb}
\usepackage{delarray,amsmath,bbm}
\usepackage[latin1]{inputenc}
\usepackage[american]{babel}
\usepackage{cite}
\usepackage{subfigure}
\usepackage{epsfig}

\newcommand{\be}{\begin{equation}}
\newcommand{\ee}{\end{equation}}
\newcommand{\beqa}{\begin{eqnarray}}
\newcommand{\eeqa}{\end{eqnarray}}
\numberwithin{equation}{section}

\newfont{\namefont}{cmr10}
\newfont{\addfont}{cmti7 scaled 1440}
\newfont{\boldmathfont}{cmbx10}
\newfont{\headfontb}{cmbx10 scaled 1728}

\begin{document}
\baselineskip=15.5pt
\pagestyle{plain}
\setcounter{page}{1}

\begin{center}
\rightline{IFT-UAM/CSIC-14-131 }
\vspace{2cm}

\renewcommand{\thefootnote}{\fnsymbol{footnote}}

\begin{center}
{\Large \bf  Collapse and Revival in Holographic Quenches }
\end{center}
\vskip 0.1truein
\begin{center}
\vspace{6mm}
\bf{Emilia da Silva${}^1$, 
Esperanza Lopez${}^1$, 
Javier Mas${}^2$ and Alexandre Serantes${}^2$ }
\end{center}
\vspace{0.5mm}

\begin{center}\it{
${}^1$ Instituto de F\'\i sica Te\'orica IFT UAM/CSIC \\
Universidad Aut\'onoma de Madrid\\
 28049 Cantoblanco, 
 Madrid, Spain
}
\end{center}

\begin{center}\it{
${}^2$Departamento de  F\'\i sica de Part\'\i  culas \\
Universidade de Santiago de Compostela, 
and \\
Instituto Galego de F\'\i sica de Altas Enerx\'\i as IGFAE\\
E-15782 Santiago de Compostela, Spain}
\end{center}

\vspace{0.5mm}

\setcounter{footnote}{0}
\renewcommand{\thefootnote}{\arabic{footnote}}

\let\thefootnote\relax\footnotetext{Emails: emilia.dasilva@csic.es, esperanza.lopez@uam.es, javier.mas@usc.es, \\ alexandre.serantes@gmail.com}

\vspace{0.6in}

\begin{abstract}
\noindent 
We study holographic models related to global quantum quenches in finite size systems. The holographic set up describes naturally  a CFT, which we consider on a circle and a sphere.
The enhanced symmetry of the conformal group on the circle motivates us to compare the evolution in both cases.
Depending on the initial conditions, the dual geometry exhibits oscillations that we holographically interpret as revivals of the initial field theory state. On the sphere, this only happens when the energy density created by the quench is small compared to the system size.
However on the circle considerably larger energy densities are compatible with revivals. Two different timescales emerge in this latter case. A collapse time, when the system appears to have dephased, and the revival time, when after rephasing the initial state is partially recovered. The ratio of these two times depends upon the initial conditions in a similar way to what is observed in some experimental setups exhibiting collapse and revivals.

\end{abstract}

\smallskip
\end{center}
\newpage


\section{Introduction}

The study of the out of equilibrium dynamics of many body isolated quantum systems embodies important open questions. On generic grounds a fast approach to ergodic behavior is expected, but there is a variety of situations where this is not realized \cite{Polkovnikov:2010yn}. Conservation laws of integrable theories constrain the final state to be different from that prescribed by the canonical ensemble \cite{Rigol2007}. There are systems which exhibit metastable intermediary states which retain memory of the initial excitation, known as pre-thermalization plateaux  \cite{Gring.et.al.2012}, and only thermalize on long time scales. The interest on this subject has been fostered by the experimental control of cold atomic systems.

Sometimes a system, initially out of equilibrium,  evolves towards an apparently decohered state which after some time exhibits a remarkable  reconstruction, termed revival in the literature. When revivals appear, they are not isolated, but repeat themselves with or without decay \cite{Robinett2004,Igloi2011,Happola2012,Cardy2014}.
The focus of this work is set on this last type of effect. Our aim is to put forward the striking similarity between  quantum revivals and  certain quasi-periodic solutions to the Einstein equations. The apparent huge conceptual  distance among these two physical phenomena is bridged by the holographic principle, whose best known realization is the AdS/CFT correspondence \cite{Maldacena:1997re}.

The holographic dictionary assigns a classical gravity system to a strong coupling field theory problem, offering in this way a radically new perspective. Out of equilibrium  processes are mapped to time dependent metrics. In particular, field theory thermalization is matched with gravitational collapse \cite{Banks:1998dd,Danielsson:1999fa}. Quantum quenches, a useful technique to generate out of equilibrium states, has been modeled holographically along these lines. One of the most salient features of the evolution after a quench is the so called light-cone effect for the propagation of quantum correlations \cite{Calabrese:2005in}. In the holographic context, this effect has been shown to be reproduced from the radial infall of a matter shell in an asymptotically anti-de Sitter geometry \cite{AbajoArrastia:2010yt,Balasubramanian:2010ce}.

Anti-de Sitter bears resemblance with a confining well. A spherical matter shell which does not carry enough energy density to create a horizon as it implodes, will scatter away.
However in an asymptotically AdS geometry, the shell will bounce back and start a new infall \cite{Pretorius:2000yu,Bizon:2011gg}. 
In  \cite{Abajo-Arrastia:2014fma} it was proposed that such bouncing geometries provide the holographic counterpart to processes which exhibit quantum revivals. 
Among the observations  that supported such interpretation a prominent one came from the time required by the shell to complete a cycle.
It was always close to the revival time predicted by the simple but successful model of \cite{Calabrese:2005in} based on the free streaming of quasiparticle excitations.
This model underlies the periodicities after a quench found in some finite size systems described by integrable spin chains \cite{Igloi2011,Happola2012}  and conformal field theories \cite{Cardy2014}.

In this note we study scalar collapse in AdS$_3$ and AdS$_4$, where the dual system corresponds to a CFT living respectively on a circle and a sphere. The infinite dimensional symmetry algebra associated with the conformal invariance on the circle is absent on the sphere. This important difference motivates the interest in performing a comparative analysis of the revivals in both cases.
Indeed we obtain a richer phenomenology on the circle, linked to comparatively larger energies being compatible with revivals. The evolution of the entanglement entropy shows that quantum correlations spread initially according to the mentioned propagation model, but the periodicites can be much longer than
expected according to it. The resulting pattern relates to a series of collapse and revivals such as found in the Jaynes-Cummings model \cite{Jaynes:1963zz}, describing a two-level atom coupled to quantized radiation in a cavity, or in some condensates of atoms trapped in optical lattices \cite{walls,Greiner2002}. Remarkably, the ratio between collapse and revival time in those examples depends upon the initial conditions in a similar way to the one we find holographically.

The plan of the paper is the following.
We review in Section 2 the holographic dictionary for scalar collapse. In Section 3 we analyze the phase diagram for bouncing geometries, trying intentionally to keep the exposition at a non-technical level. The evolution of the entanglement entropy is addressed in Section 4, and Section 5 is devoted to the physical interpretation of our results in terms of the dual field theory. A comparison with known systems exhibiting collapse and revivals is performed in Section 6. Section 7 contains a qualitative map between the initial profile of the imploding matter shell and the entanglement properties of the dual field theory state.
Finally, we address some open questions and summarize our conclusions in Section 8.

\section{Bouncing geometries}

The holographic dictionary relates classical gravity in an AdS$_{d+1}$ space with the vacuum of a CFT$_d$ at strong coupling and large central charge \cite{Maldacena:1997re}. Black holes in AdS are interpreted as the dual geometry for thermal states of the dual CFT \cite{Witten:1998zw}. Along this line, the gravitational collapse of a certain matter configuration to form a black hole is considered the holographic representation of a relaxation process on the field theory side \cite{Banks:1998dd,Danielsson:1999fa}. 

We wish to study the relaxation dynamics of finite size systems. We focus on field theories living on $(d\!-\!1)$-dimensional spheres, since this is easiest to implement holographically.  
For simplicity we also restrict to homogeneous field theory states, as are those implied in global quenches. Hence we are led to study gravitational collapse of a spherically symmetric matter distribution. A convenient ansatz for the metric is \cite{Bizon:2011gg} 
\be
ds^2 = \frac{1}{\cos^2 x}\left( - A(t,x) e^{-2\delta(t,x)} dt^2 + {dx^2 \over A(t,x)} + \sin^2 x\, d\Omega^2_{d-1}\right) \, ,
\label{metric}
\ee
where $d\Omega_{d-1}^2$ is the line element of a unit $(d\!-\!1)$-sphere, $x\in[0,\pi/2]$ and the AdS radius have been set to one. The static geometry
\be
A(t,x)=1\, , \hspace{1cm}  \delta(t,x)=0 \, ,
\label{ads}
\ee
describes pure AdS.

\begin{figure}[h]
\begin{center}
\includegraphics[width=7.8cm]{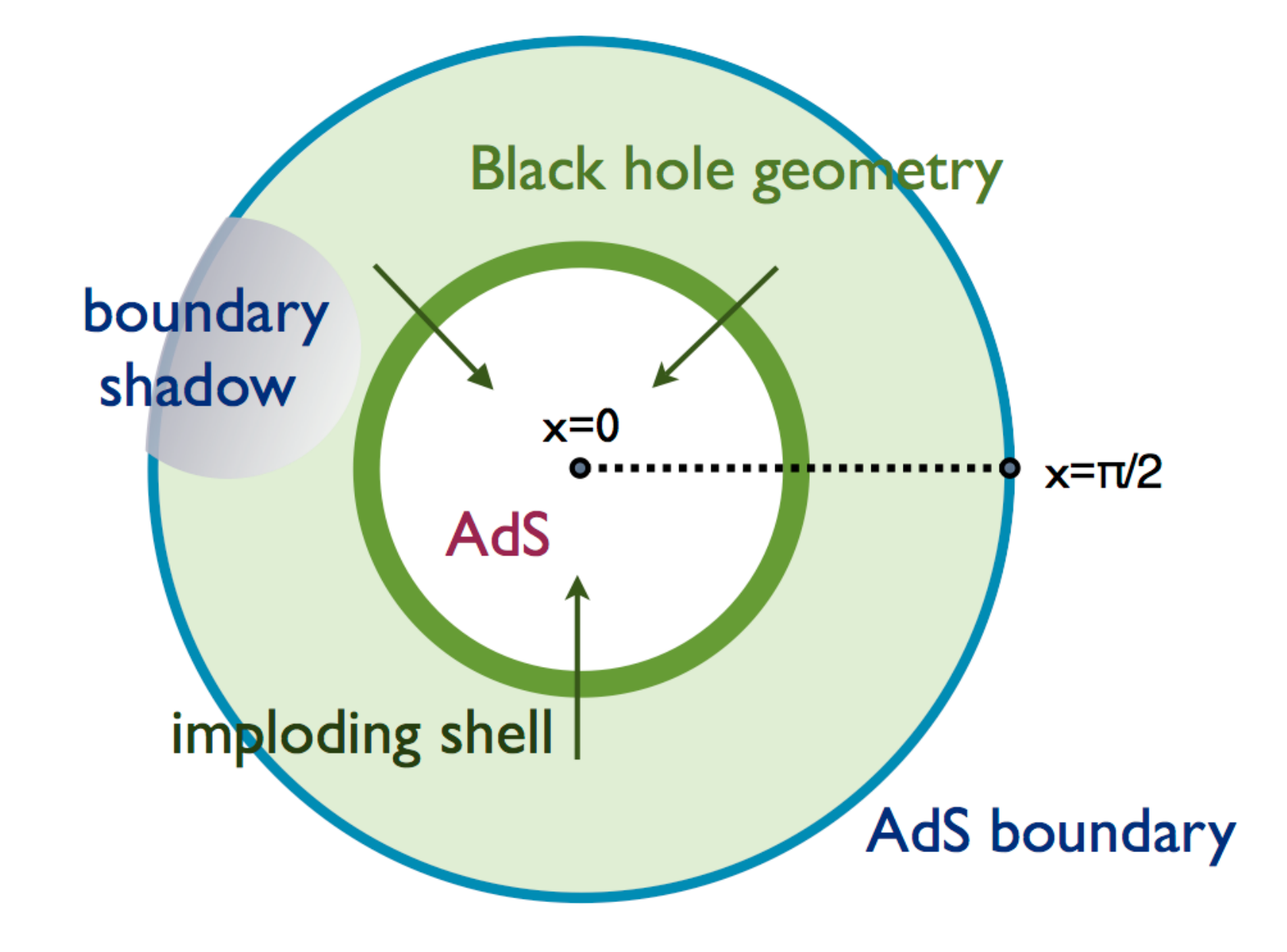}
\vspace{-3mm}
\end{center}
\caption{\label{fig:picture} Collapse of a matter shell in AdS. Outside the shell the geometry is that of a black hole, inside is empty AdS. We have represented by a shadowed area the bulk region implied in the holographic computation of observables for a certain boundary region. It reaches further in the interior the larger the boundary region under consideration.}
\end{figure}

The radial coordinate $x$ is to be interpreted as a scale for the dual field theory, $x\!=\!0(\pi/2)$ encoding its long(short) distance physics. Gravity induces any radial distribution of matter to start imploding. The geometry outside the matter shell is that of a black hole of the same total mass. Hence as the shell infalls, field theory observables associated to larger scales produce thermal results, see Fig.\ref{fig:picture}. This has been shown to successfully reproduce the horizon effect characteristic of the evolution after quantum quenches \cite{AbajoArrastia:2010yt,Balasubramanian:2010ce}, leading to the following interpretation. In the collapse processes we are considering, $x$ acts as a scale measuring the typical separation of entangled excitations in the field theory out of equilibrium state \cite{Abajo-Arrastia:2014fma}. A shell sharply localized close to the AdS boundary represents entanglement mainly among nearest neighbors. The shell implosion is the holographic counterpart to the flight apart of entangled excitations.

\begin{figure}[h]
\begin{center}
\includegraphics[width=15cm]{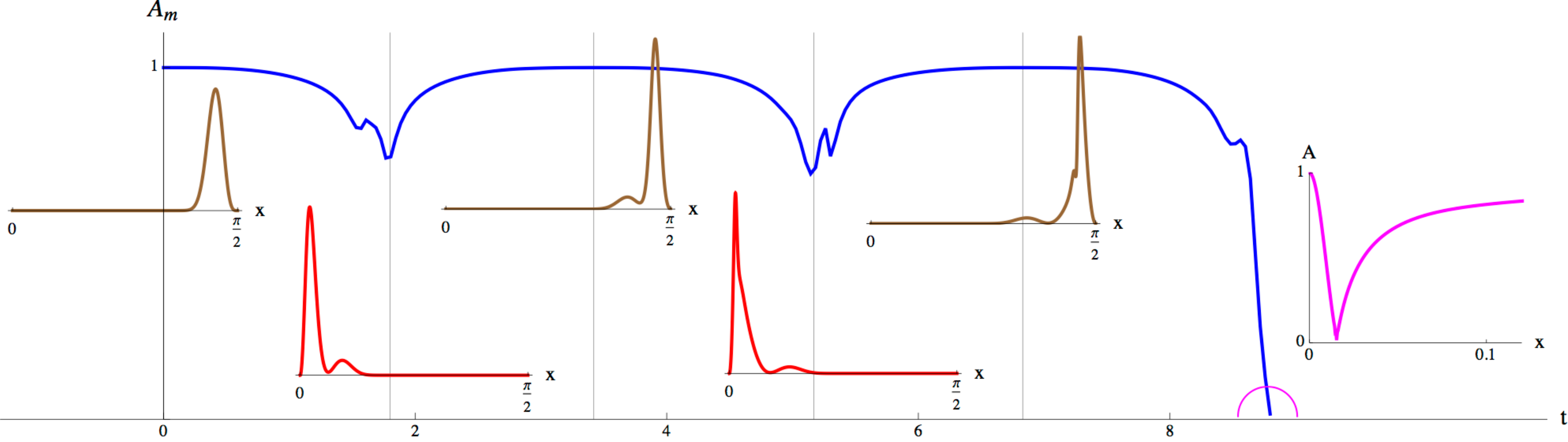}
\vspace{-3mm}
\end{center}
\caption{\label{fig:F} In blue, evolution of $A_m(t)$ for a typical thin shell in AdS$_4$. It needs two bounces with the AdS boundary for collapse. In brown and red, the radial energy distribution of the shell at the times indicated by the vertical lines. In magenta, $A$ develops a zero signaling black hole formation.}
\end{figure}

If the matter shell is not sufficiently localized in the radial direction as it falls, or it does not carry enough mass, it implodes, scatters against itself and starts expanding without forming a horizon. When reflecting conditions are imposed at the AdS boundary, $x\!=\!\pi/2$, the shell will subsequently bounce against it and start a new implosion \cite{Pretorius:2000yu,Bizon:2011gg}. Such cycle can repeat several times. It was proposed in \cite{Abajo-Arrastia:2014fma} that these bouncing oscillations serve as the holographic duals for revivals found in condensed matter systems. As a way of characterizing the revivals we will use the function $A(t,x)$ in \eqref{metric}. When the matter shell is close to the AdS boundary, the inverse cosine factor multiplying the {\it rhs} of \eqref{metric} dilutes its density and minimizes its back reaction on the geometry. The resulting geometry is close to pure AdS, namely $A(t,x)\!\approx\!1$. In a black hole geometry, $A(t,x)$ vanishes at the event horizon. In a dynamical situation, a zero of $A(t,x)$ signals the presence of trapped surfaces that capture part of the shell mass. Holographically this represents the irreversible dephasing of some field theory degrees of freedom, or equivalently, the onset of equilibration. Hence
\be
A_m(t)=\min_{x} A(t,x)
\label{F}
\ee
can be used as a measurement of how far from equilibration the system is. This function takes values on the interval $[0,1]$. The minima of $A_m(t)$ correspond to the moments of maximal implosion of the matter shell and the maxima to its bounces against the AdS boundary, see Fig.\ref{fig:F}. 

Following previous works \cite{Bizon:2011gg,Jalmuzna:2011qw,Garfinkle:2011hm,Garfinkle:2011tc,Buchel:2012uh,Buchel:2013uba,Bizon:2013xha}, we will consider a matter shell composed of a massless scalar field. The equations of motion for the system are \cite{Bizon:2011gg}
\beqa
\dot\Phi &=& \left( A e^{-\delta} \Pi\right)' ~~~,~~~\dot \Pi = \frac{1}{\tan^{d-1} x}\left(\tan^{d-1} x\, A e^{-\delta} \Phi\right)' \, ,\label{eqforphi}
\\
A' &=& {d-2+2 \sin^2 x \over \sin x \cos x} (1-A) -\sin x \cos x \, A\,  (\Phi^2 + \Pi^2)\, ,\label{eqforA}\\[2mm]
\delta' &=& -\sin x \cos x \,(\Phi^2 + \Pi^2)\, . \label{eqford}
\eeqa
where $\Phi\!=\!\phi'$ and $\Pi\!=\!A^{-1} e^\delta \dot{\phi}$, with $\phi'$ and  $\dot{\phi}$ the space and time derivatives of the scalar field respectively. For convenience, the scalar field has been chosen to have an unconventional normalization 
\be
\phi=\sqrt{8 \pi G \over d-1}\, \varphi \, , 
\label{resc}
\ee
where $\varphi$ is a canonically normalized scalar field and $G$ the Newton's constant.

We solve the previous system of equations by setting the profile of the scalar field derivatives at $t\!=\!0$. We will use a gaussian type profile located close to $x\!=\!\pi/2$  as initial data \cite{Abajo-Arrastia:2014fma}
\be
\Phi (0,x)=0\, , \hspace{1cm} \Pi (0,x)=\epsilon   \exp\left({-{4 \tan^2 (\pi/2 - x) \over \pi^2 \sigma^2}}\right) \cos^{d-1} x\, .
\label{profile}
\ee
The parameter $\sigma$ controls the thickness of the shell, and and for given $\sigma$ its mass is a monotonous function of $\epsilon$. Values $\sigma\!<\!0.1$ corresponds to thin shells sharply localized at the boundary, while $\sigma\!>\!0.5$ describes broad initial shells with evolve into radially delocalized pulses as they fall. In Section 7 we will argue that the shell thickness is holographically related to the time span of the perturbation creating the field theory out of equilibrium state \cite{Abajo-Arrastia:2014fma}. Very thin pulses correspond to instantaneous actions on the field theory while broad profiles are associated with slow perturbations. 

The initial data \eqref{profile} have to be supplemented with additional boundary conditions. A holographic model for a closed field theory system is obtained imposing  
the absence of energy exchange throughout the AdS boundary. Under this condition the mass of the matter distribution keeps constant along the evolution. Recalling the normalization of the scalar field \eqref{resc}, it is  given by
\be
M={(d-1)\over 16 \pi G}\;  {\rm vol}(S^{d-1}) \, {\cal M} \, ,
\label{mass0}
\ee
with ${\rm vol}(S^{d-1})$ the volume of a unit $(d\!-\!1)$-sphere and
\be
{\cal M}=\int_0^{\pi/2}  (\tan x)^{d-1}  (\Phi^2+\Pi^2) A  \, dx \, .
\label{mass}
\ee
The shell mass is identified with the energy of the dual CFT. Its central charge, which grows with the number of elementary degrees of freedom, holographically relates to the inverse of the Newton's constant \cite{Brown:1986nw}. Hence $\cal M$ provides a measurement of the energy density per species in the field theory.

The system of equation \eqref{eqforphi}-\eqref{eqford} includes an integration constant related to the presence and strength of a curvature singularity at $x\!=\!0$. We fix it by requiring the smoothness of the geometry at the origin of the radial coordinate \cite{Bizon:2011gg}. With this choice, and although  the generic result of the gravitational dynamics is the formation of a black hole, it can be explicitly shown that holographic set up describes a unitary field theory evolution from an initial pure state \cite{AbajoArrastia:2010yt,Takayanagi:2010wp}. Hence the gravitational collapse of an initial shell \eqref{profile} with small parameter $\sigma$ can be used as a reasonable model for the evolution after a sudden global quench. 

The metric \eqref{metric} is invariant under reparameterizations of the time direction $t\!\rightarrow\!t'\!=\!f(t)$. Since we aim at the holographic interpretation of the geometry, we ask $t$ to be the proper time on the AdS boundary. Namely the metric on the asymptotic slice $x\!\rightarrow\!\pi/2$ should be conformally equivalent to 
\be
ds^2= -  dt^2 + d\Omega^2_{d-1} \, ,
\ee
which is the natural field theory metric. This fixes the previous gauge freedom and completes the set of initial and boundary data. See \cite{Bizon:2011gg} for further technical details on the initial value problem. The numerical integration of \eqref{eqforphi}-\eqref{eqford} has been accomplished using a fourth-order Runge-Kutta algorithm. We use below the mass of the shell and its thickness, codified in ${\cal M}$ and $\sigma$, as physical parameters to perform a scan over the initial conditions \eqref{profile}.

\section{Revival time}

We are interested in gravitational processes where a mater shell does not generate a trapped surface by direct collapse, but needs at least one bounce off the AdS boundary to do so. Different aspects of scalar collapse in AdS$_{d+1}$ have been studied in \cite{Bizon:2011gg,Jalmuzna:2011qw,Garfinkle:2011hm,Garfinkle:2011tc,Buchel:2012uh,Buchel:2013uba,Bizon:2013xha}. We will focus here on how the time invested on the first bounce depends on the mass and thickness of the initial shell. As argued above, we interpret this time as that at which the initial out of equilibrium state of the dual CFT undergoes the first revival. 

An interesting question is how symmetry constraints influence the evolution of a system towards relaxation. With this motivation in mind, we will compare the phenomenology of scalar collapse in AdS$_3$ and AdS$_4$, dual respectively to a CFT on a circle and a sphere.

There are important differences between AdS$_{3,4}$ gravity already at the level of static solutions: in AdS$_3$ there is a mass threshold for the existence of black holes \cite{Banados:1992wn}. For the choice of unit circle made in \eqref{metric}, curvature singularities are hidden behind a horizon only for masses above
\be
{\cal M}  =1 \, .
\label{thres}
\ee
Static geometries with mass below threshold contain an unshielded conical singularity at their center. On the contrary,  AdS$_4$ admits black holes of any positive mass.

\begin{figure}[h]
\begin{center}
\includegraphics[width=7.6cm]{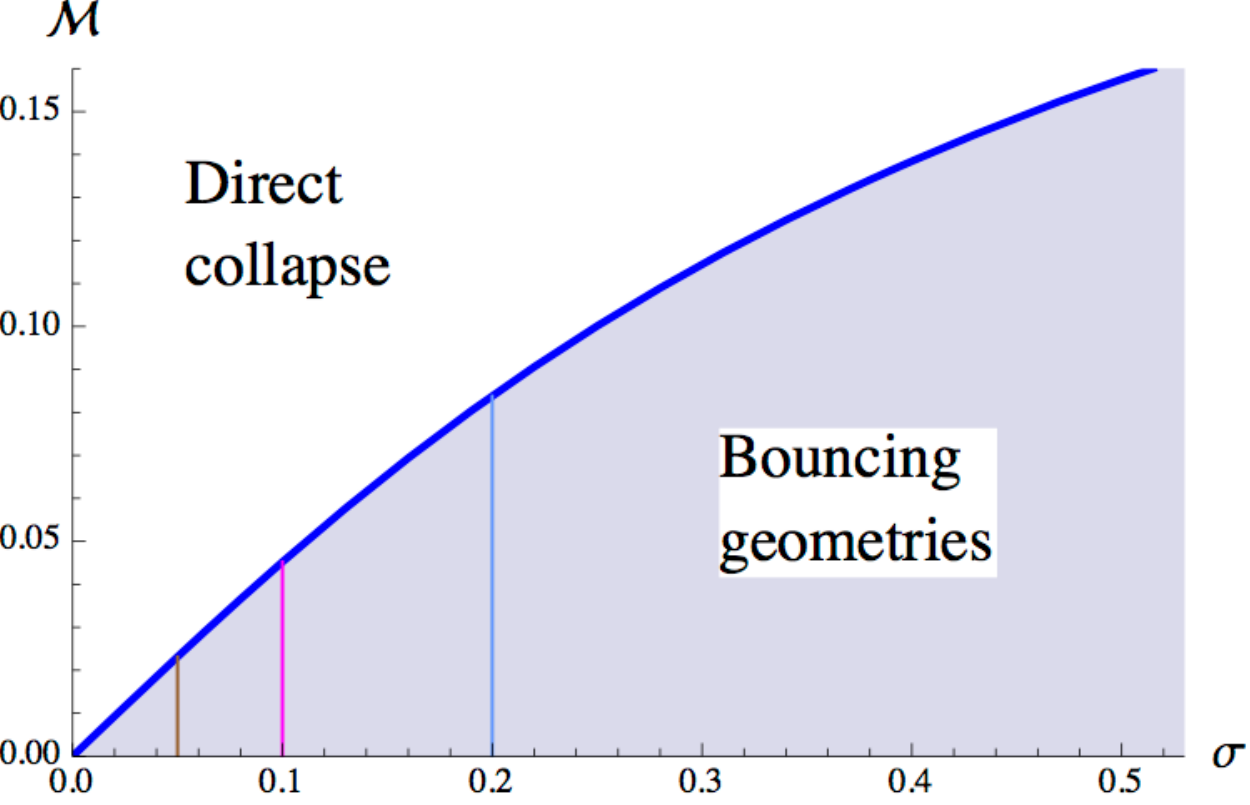}~
\includegraphics[width=7.9cm]{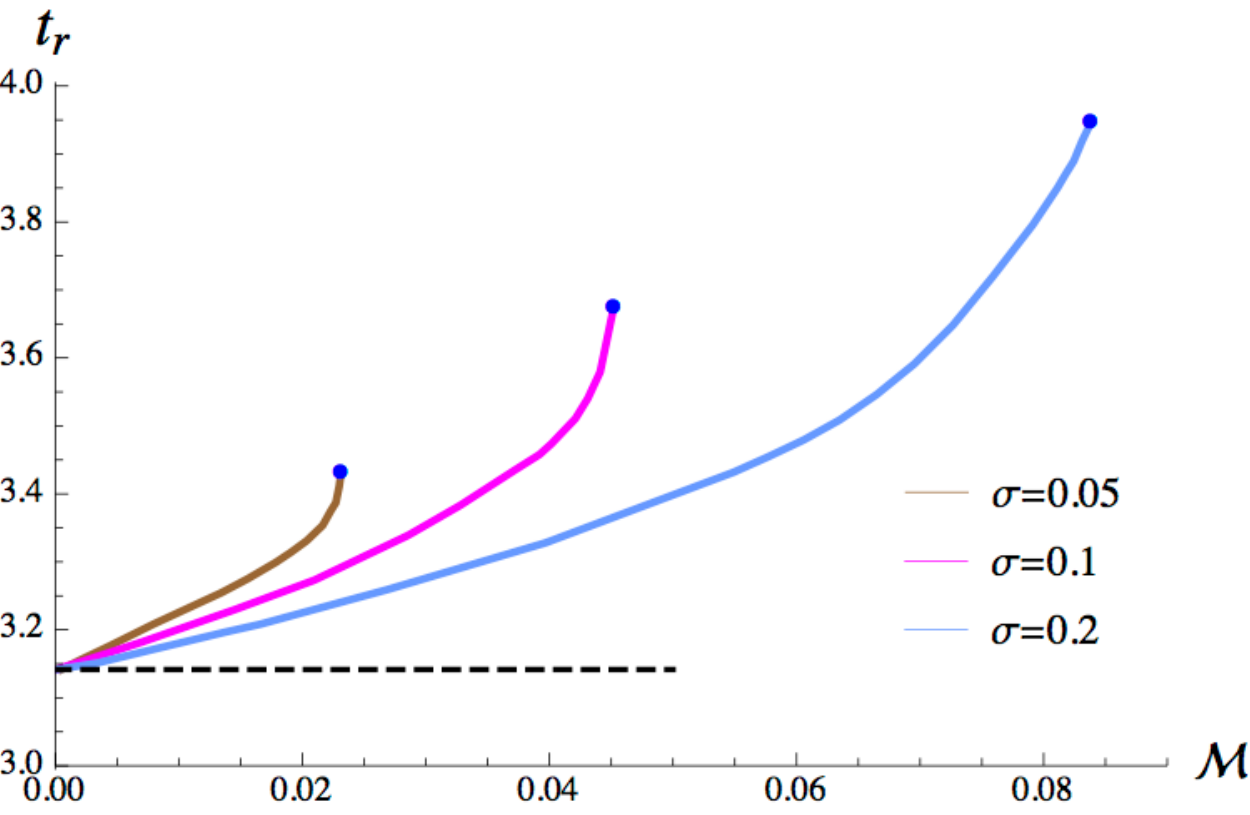}
\end{center}
\caption{\label{fig:AdS4} Left: Phase diagram for scalar collapse with initial data \eqref{profile} in AdS$_4$. The shaded region signals processes requiring at least one bounce for collapse. Right: Dependence of the bouncing period with the mass for fixed $\sigma\!=\!0.05,0.1,0.2$. The blue dots signal the threshold value for direct collapse. The dashed line is $\tau\!=\!\pi$.}
\end{figure}

This has important consequences for scalar collapse. In AdS$_4$ any thin shell appears to induce the formation of a black hole after sufficient number of bounces off the boundary \cite{Bizon:2011gg}. The number of bounces required for the emergence of trapped surfaces decreases with increasing mass\footnote{When a trapped surface emerges, it generically does not capture the complete shell. A fraction of it can yet escape to the boundary and require several further bouncing cycles to be completely absorbed. See \cite{Abajo-Arrastia:2014fma} for an analysis of the subsequent evolution.}, until above some value this happens at the first implosion.  The limiting line among these two situations is depicted in Fig.\ref{fig:AdS4}a. We observe that bouncing geometries are only obtained for quite small masses \cite{Bizon:2011gg,Buchel:2013uba,Abajo-Arrastia:2014fma}. Correspondingly, revivals in the dual CFT$_3$ happen for initial out of equilibrium states with an energy density per species clearly small compared to the system size.

\begin{figure}[h]
\begin{center}
\includegraphics[width=7.5cm]{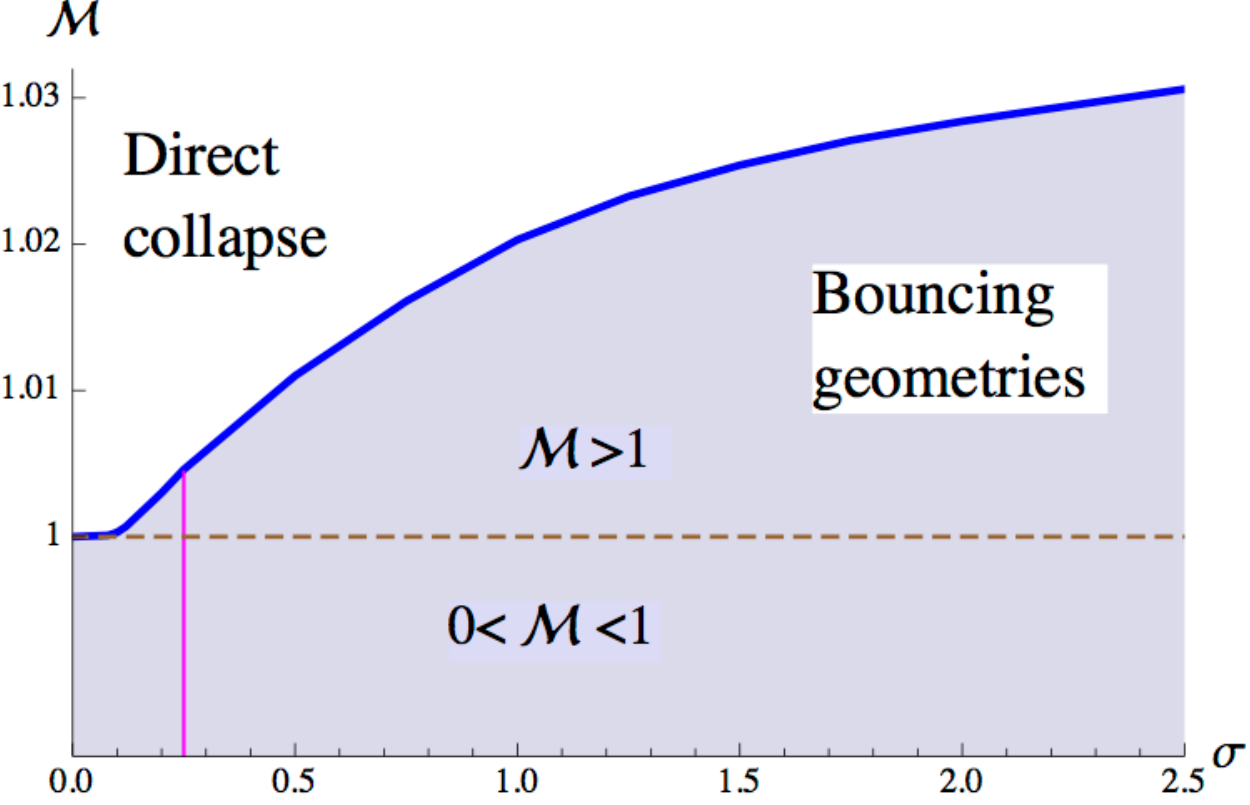}~
\includegraphics[width=7.5cm]{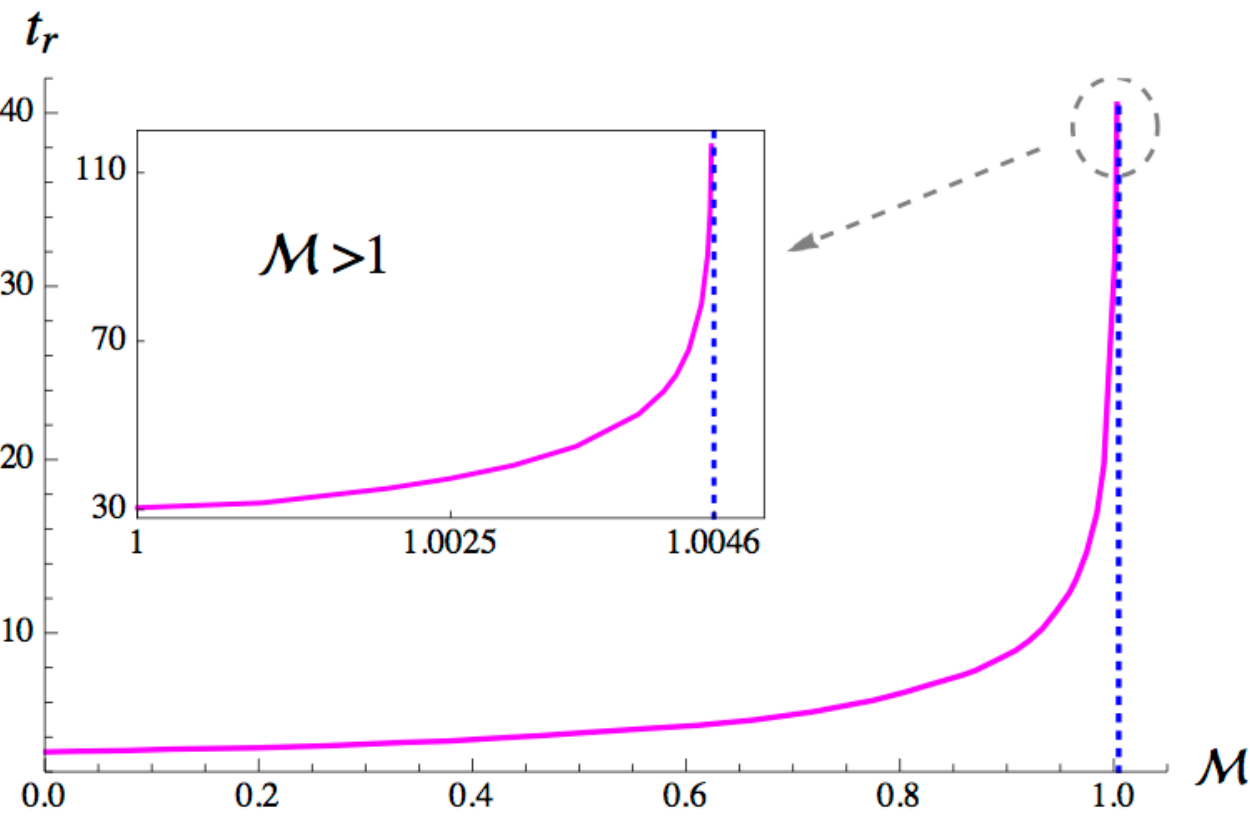}
\end{center}
\caption{\label{fig:period} Left:  Phase diagram for scalar collapse with initial data \eqref{profile} in AdS$_3$. Right: Mass dependence of $t_r$ for initial pulses \eqref{profile} with $\sigma\!=\!0.25$ in AdS$_3$. In the inset, detail of the plot in the small window above threshold compatible with bounces. The apparent change of slope is due to the rescaling of the vertical axes.}
\end{figure}

In contrast to AdS$_4$, the eventual collapse of shells below the threshold \eqref{thres} in AdS$_3$  could only end up forming a naked singularity. As far as we could push our simulations and keep them under numerical control, we have not found this to be the case. For low masses, the singularity analysis of \cite{Bizon:2013xha} points towards excluding the formation of such singularity in finite time. In these cases equilibration is never achieved and instead the dual field theory undergoes an infinite series of revivals. 
Even above the threshold \eqref{thres}, there is a small window where at least one bounce off the AdS$_3$ boundary is required before collapse \cite{Pretorius:2000yu}. Fig.\ref{fig:period}a shows the curve separating bouncing geometries in AdS$_3$ from trapped surface formation at the first implosion.
Hence holographic models of CFT$_2$ dynamics admit revivals for ratios of the energy density to the system size much larger than in higher dimensions.

For concreteness, we characterize the revival time as the time invested in completing the corresponding oscillation cycle of the function $A_m(t)$ in \eqref{F}, see Fig.\ref{fig:F}. 
Generically the shell motion is quasi-periodic, and the revival time stays almost constant along the evolution. However for AdS$_3$ shells close or above the black hole threshold, ${\cal M}\!\gtrsim\!1$,  the time span of successive bounces might vary.
Therefore, to avoid ambiguities, we will denote with $t_r$  the time  elapsed in the first revival. Both in AdS$_3$ and AdS$_4$, the value of $t_r$ tends to $\pi$ for scalar pulses of low mass, ${\cal M}\!\to\!0$.
This is expected since a null ray originating at the boundary and traversing diametrically AdS returns to the boundary after a time $t\!=\!\pi$, as can be easily obtained from \eqref{metric} and \eqref{ads}. The value of $t_r$ monotonically increases with the shell mass, see Fig.\ref{fig:AdS4}b and \ref{fig:period}b.
 
The mass window for the existence of bouncing geometries in AdS$_4$ closes down to zero as the thickness of the shell vanishes $\sigma\!\to\!0$, see Fig.\ref{fig:AdS4}a. For this reason the revival time associated to thin pulses in AdS$_4$ is always approximately $\pi$. Pulses of intermediate broadness need small but finite masses for direct collapse. They can reach revival times that, although close, are appreciably larger than $\pi$. This phenomenology is illustrated in Fig.\ref{fig:AdS4}b.

The comparatively higher masses compatible with revivals in AdS$_3$ have a drastic impact on the allowed values of $t_r$. The revival time strongly increases for shells whose mass approaches ${\cal M}\!=\!1$, see Fig.\ref{fig:period}b. The increase is entirely due to the shell being kept by its own gravitational potential at the point of maximal implosion  for a long time before expanding again. Pulses leading to revivals with $\cal M$ higher than one, take extremely long to complete the first bounce. Our results suggest that $t_r$ possibly diverges at the upper end of the mass window compatible with bounces. This is shown in the inset of Fig.\ref{fig:period}b. The relevance of the mass window above the black hole threshold relies on providing transition processes between infinite revivals and fast thermalization in holographic CFT$_2$ models. Consistently with the AdS$_4$ results, it closes down for thin shells.

When AdS$_3$ pulses with ${\cal M}\!>\!1$ implode without collapsing, they develop several extremely narrow spikes. Small differences in the initial pulse get amplified in this process and have an strong influence in the subsequent evolution. As a result the periodicity of successive bounces shows a random behavior, a feature not observed in low mass pulses. While the time required to complete the first bounce exhibits a regular monotonic growth with $\cal M$, that of subsequent bounces can both increase or decrease with the shell mass. Also the number of bounces for collapse does not decrease smoothly with the mass: for the $\sigma\!=\!0.25$ profiles reported in Fig.\ref{fig:period}a, ${\cal M}=1.004$ appears to need only one bounce for collapse while ${\cal M}=1.0045$ requires several. The described features point towards a role of chaos in the evolution of these AdS$_3$ pulses, a fact that was already noticed in \cite{Pretorius:2000yu}.

\begin{figure}[h]
\begin{center}
\includegraphics[width=7.3cm]{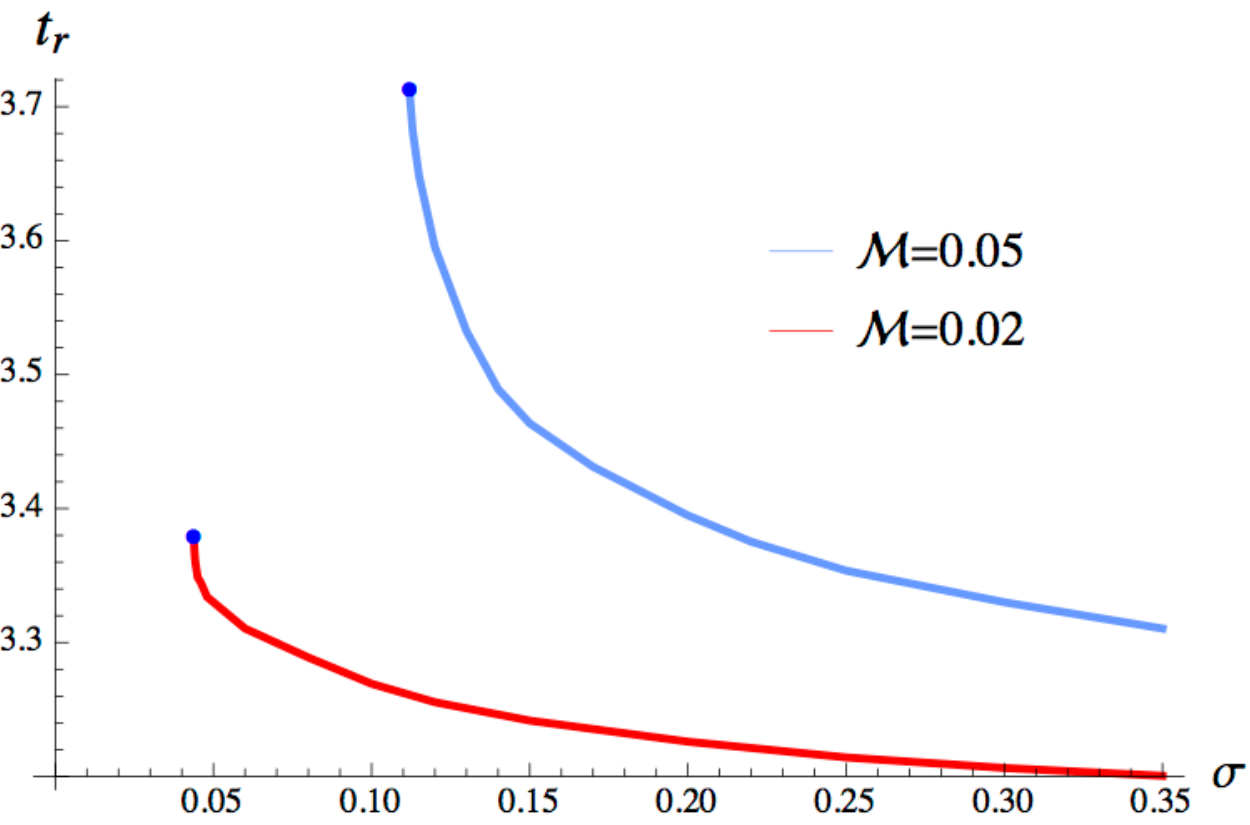}~
\includegraphics[width=7.3cm]{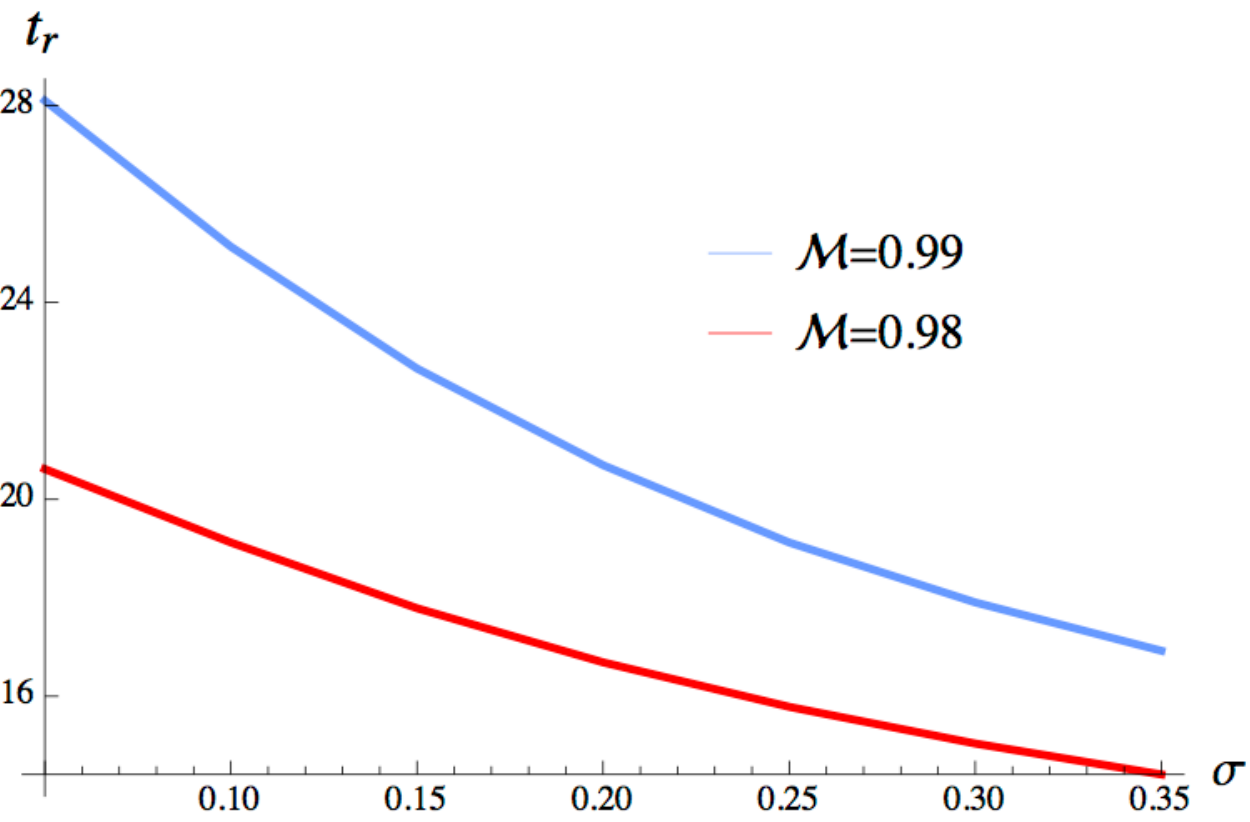}
\end{center}
\caption{\label{fig:sigma} Left: Variation of $t_r$ with $\sigma$ 
for AdS$_4$ pulses of fixed mass ${\cal M}\!=\!0.02,0.05$. The blue dots signal the threshold value for direct collapse. Right: Variation of $t_r$ with $\sigma$ 
for AdS$_3$ pulses of fixed mass ${\cal M}\!=\!0.98,0.99$.}
\end{figure}

Finally, it is interesting to notice that the time span of the first revival not only behaves smoothly as a function of the shell mass but also of its thickness.
Indeed $t_r$ decreases monotonically with $\sigma$, as shown in Fig.\ref{fig:sigma}a and Fig.\ref{fig:sigma}b for AdS$_3$ and 
AdS$_4$ pulses respectively.

\section{Entanglement entropy}

The construction of a detailed dictionary between the gravitational dynamics and the field theory evolution is far from straightforward. In the collapse backgrounds there is no time-like Killing vector that could extend field theory constant time slices into the higher dimensional dual geometry. The description of the field theory time evolution should be based on the evaluation of holographic observables. To this aim we choose the entanglement entropy, as done in many previous works.

The entanglement entropy (EE) of a region $A$ is defined as the von Neumann entropy of the density matrix traced over the degrees of freedom outside that region. This very interesting quantity, which provides a characterization of entanglement properties in extended systems, is generally difficult to calculate within field theory methods. However it has a rather accessible holographic representation. It is encoded in the area of the extremal bulk surface $\gamma_A$ that anchors at the AdS boundary on the boundary of A \cite{Ryu:2006bv,Hubeny:2007xt}
\be
S_A={\rm{Area}(\gamma_A) \over 4G} \, .
\label{HEE}
\ee

The area of bulk surfaces anchoring on the AdS boundary receives a divergent contribution from the region close to the boundary. Since the geometries we are studying
 are asymptotically AdS as $x\!\rightarrow\!\pi/2$, we can obtain a finite quantity by substituting the numerator in \eqref{HEE} by 
\be
L_A= {\rm Area} (\gamma_A)-{\rm Area}(\gamma_A^{AdS}) \, ,
\label{EER}
\ee
where $\gamma_A^{AdS}$ is the extremal surface in pure AdS ending on the same boundary region. This is equivalent to eliminating the intrinsic cutoff dependence of the entanglement entropy by defining it with respect to a reference value, which we take as that in the CFT vacuum.  

\begin{figure}[h]
\begin{center}
\includegraphics[width=7.8cm]{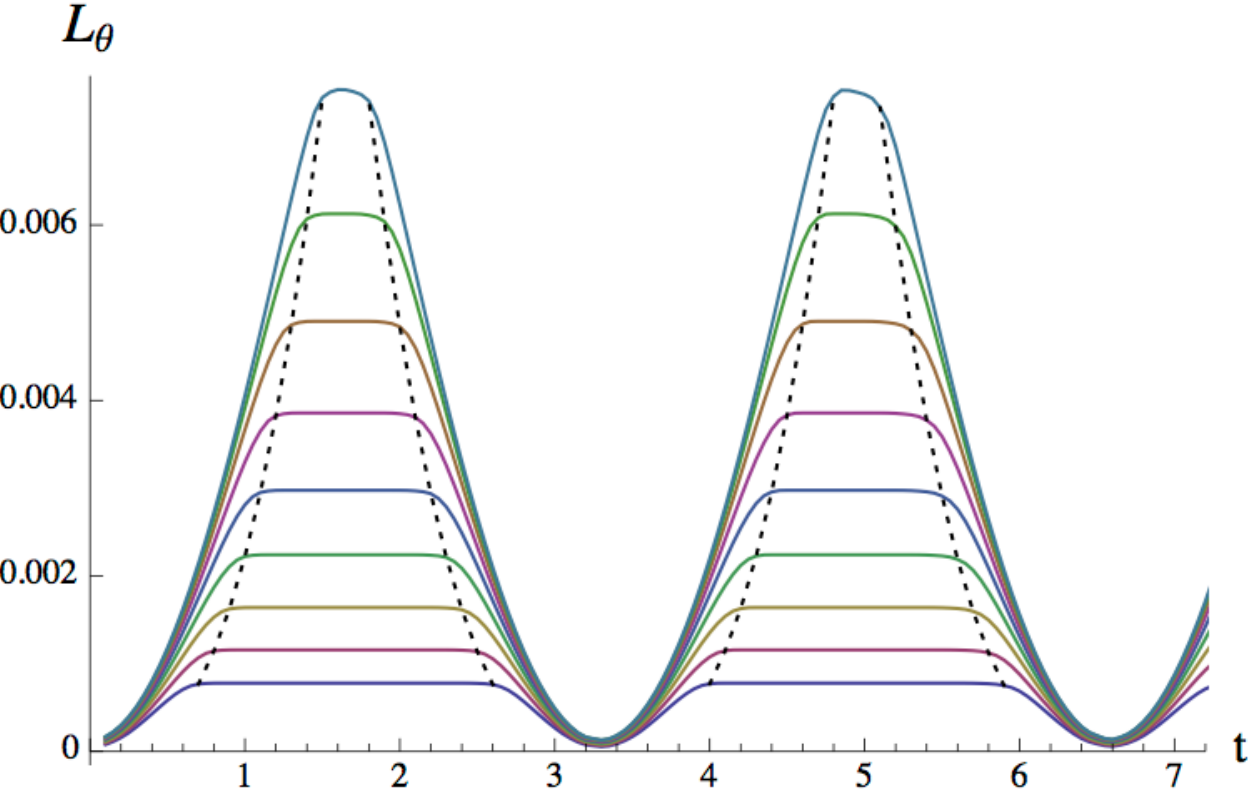}
\includegraphics[width=7.8cm]{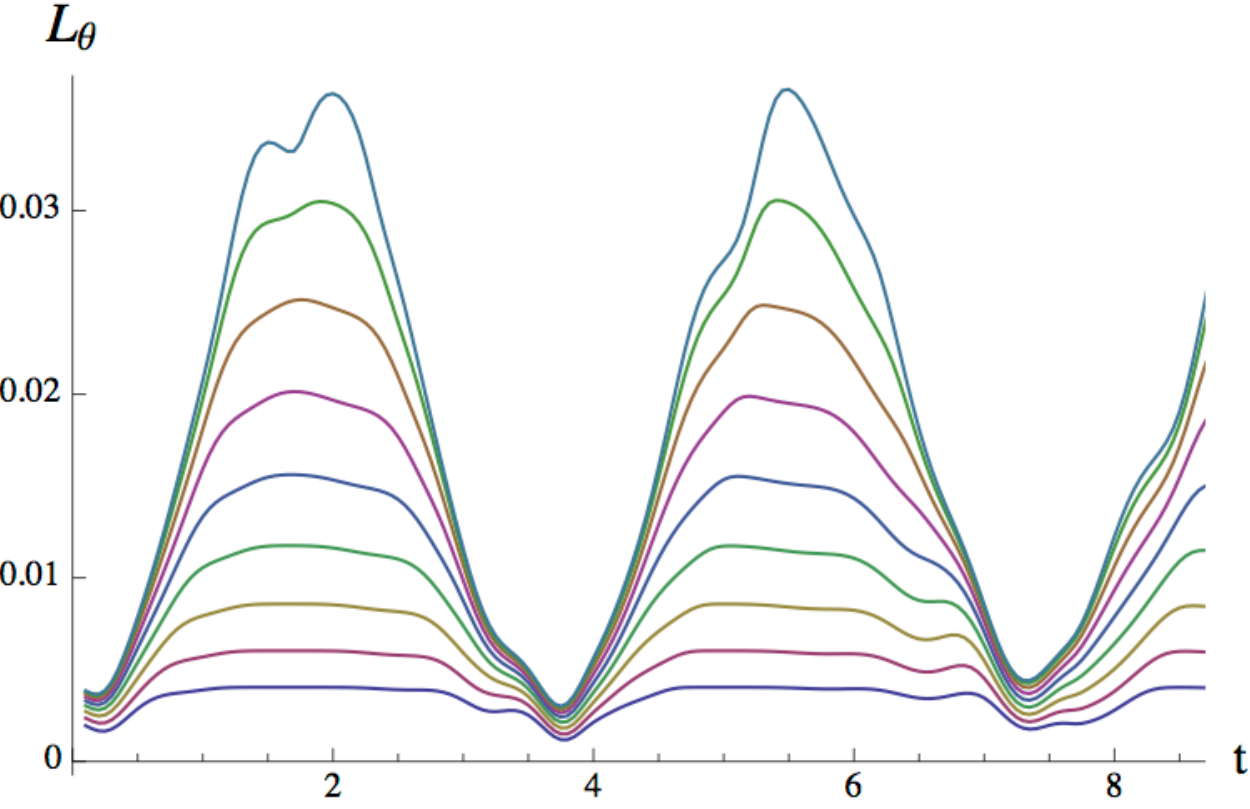}
\end{center}
\caption{\label{fig:EE4} EE evolution of spherical caps with $\theta\!=\!1.4,..,3$ in two AdS$_4$ processes which bounce twice before collapse. Left: Scalar profile with $\sigma\!=\!0.05$ and ${\cal M}\!=\!0.017$. The dotted lines signal $t\!=\!\theta/2, t_r\!-\!\theta/2, t_r\!+\!\theta/2, 2t_r\!-\!\theta/2$. Rigth: $\sigma\!=\!0.3$ and ${\cal M}\!=\!0.09$. }
\end{figure}

Processes in the CFT$_3$ dual to AdS$_4$ evolve on the unit two-sphere. The most symmetric choice for the region $A$ are then spherical caps. We parameterize them in terms of their angular aperture $\theta\!\in\![0,\pi]$, with $\theta\!=\!\pi$ corresponding to a hemisphere. Thin shells located initially close to the AdS boundary gave rise to the following evolution pattern in entanglement entropy of spherical caps. The backreaction on the geometry of a shell placed closed to the boundary is small, as explained in section 2, and the resulting geometry is approximately pure AdS. With the regularization \eqref{EER}, the value of the entanglement entropy at $t\!=\!0$ is almost vanishing. As the shell falls, the entanglement entropy grows. It achieves a maximum at $t\!\approx\!\theta/2$, after which its associated extremal surface keeps outside the imploding shell \cite{Abajo-Arrastia:2014fma}.  The value at the maximum coincides with that of the EE of a cap in a AdS black hole of the same total mass. If the shell does not form a trapped surface by direct collapse, it will start expanding. At $t\!\approx\!t_r\!-\!\theta/2$ it intersects the EE extremal surface again. From then on the entanglement entropy decreases until it approximately vanishes at $t\!=\!t_r$. This behavior repeats in subsequent bounces, see Fig.\ref{fig:EE4}a. We have marked with dotted lines $t\!=\!\theta/2$ and $t\!=\!t_r\!-\!\theta/2$ in the first bounce, and $t\!=\!t_r\!+\!\theta/2$ and $t\!=\!2t_r\!-\!\theta/2$ in the second, showing the accuracy of the described pattern.

In AdS$_4$ only thin shells of low mass avoid direct collapse, and for them $t_r\!\approx\!\pi$. In order to have processes with $t_r$ appreciably larger than $\pi$, it is necessary to consider broader shells. But for them the simple oscillation pattern of the entanglement entropy described above does not hold. The transition at $t\!\approx\!\theta/2$ and $t\!\approx\!t_r\!-\!\theta/2$ is not sharp, neither the value of the entanglement entropy exhibits a well defined plateau at its maximum as shown in Fig.\ref{fig:EE4}b. 
We will argue below that broad shells are holographically related to excited states created by a field theory action with a finite time span, rendering natural a more involved evolution of entanglement.

\begin{figure}[h]
\begin{center}
\includegraphics[width=7.8cm]{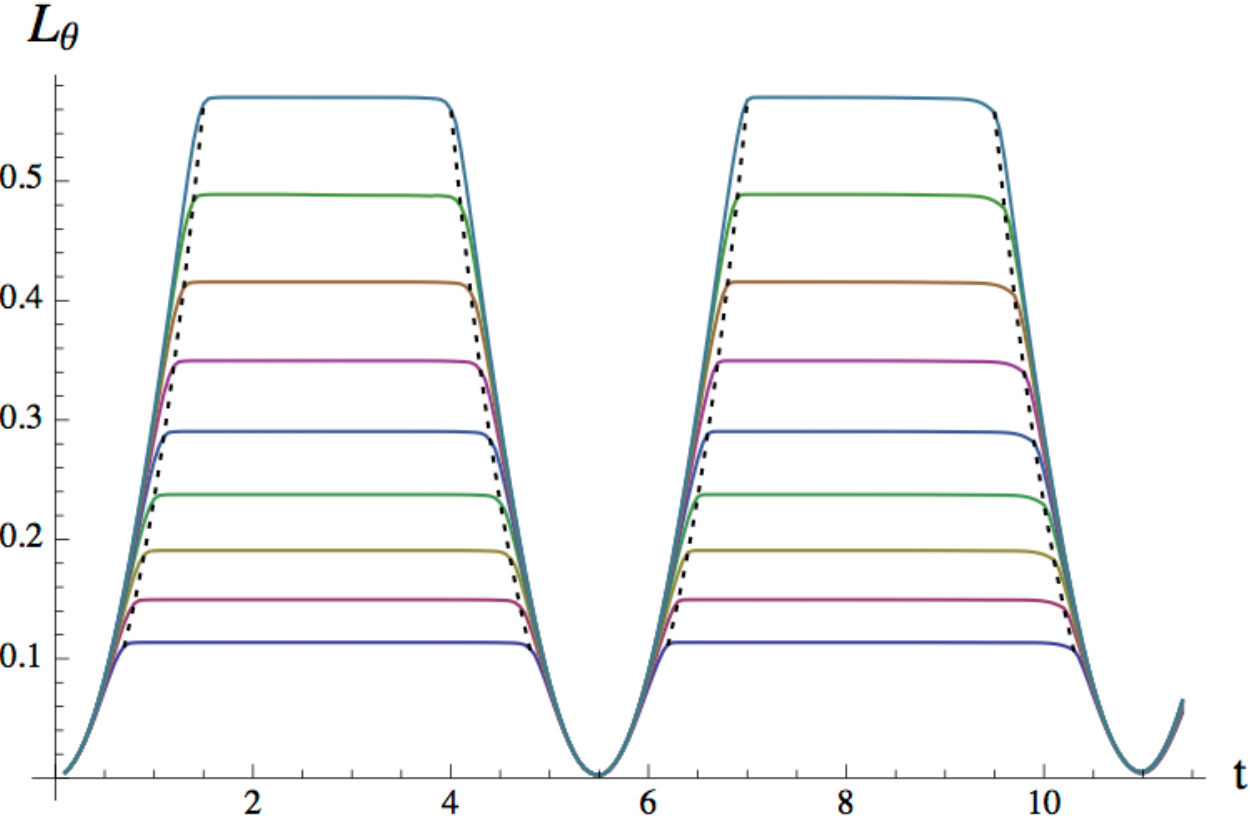}~
\includegraphics[width=7.8cm]{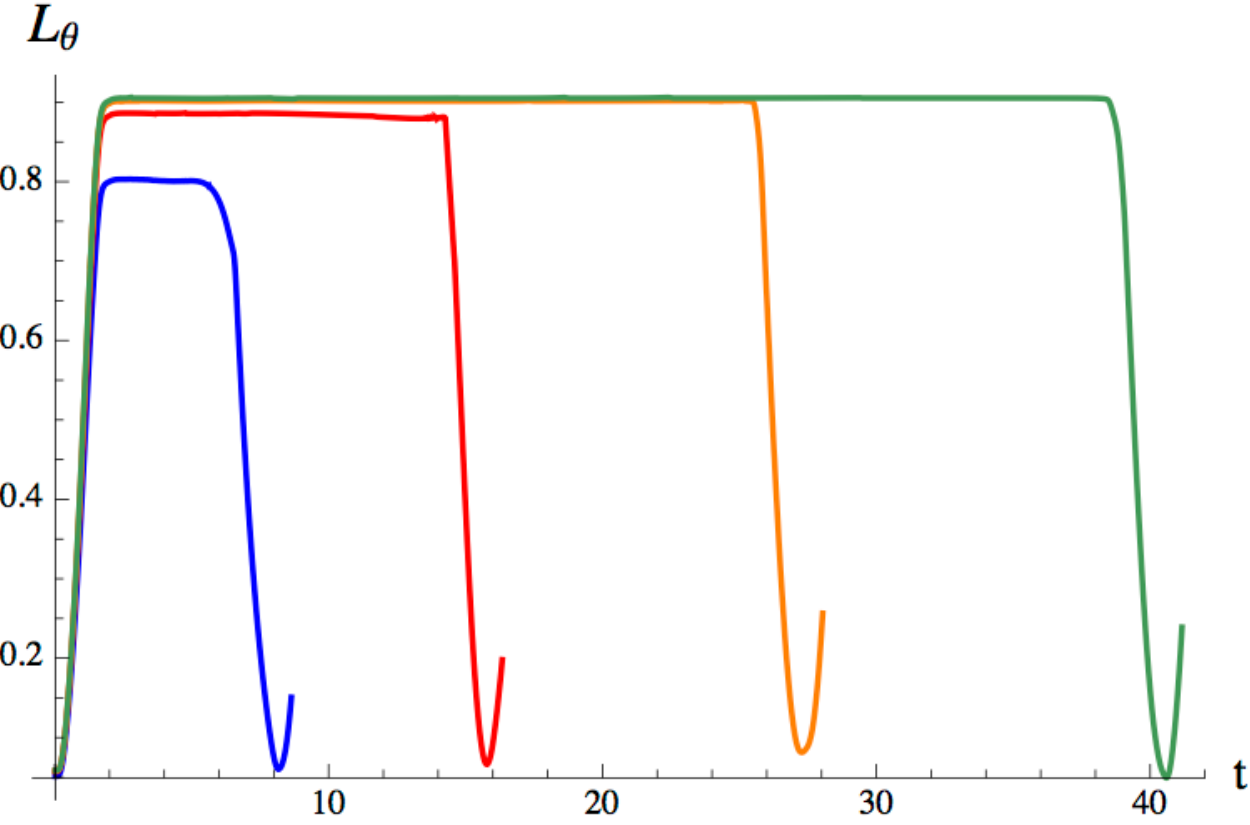}
\end{center}
\caption{\label{fig:EE3} Left: Same as in Fig.\ref{fig:EE4}a,b for an AdS$_3$ shell with $\sigma\!=\!0.05$ and ${\cal M}\!=\!0.68$. Right: Entanglement entropy of an interval with $\theta\!=\!3.14$, almost a semicircle, along the first bouncing cycle of AdS$_3$ shells with $\sigma\!=\!0.25$ and ${\cal M}\!=\!0.88,0.98,1,1.003$.}
\end{figure}

AdS$_3$ offers the possibility to explore processes generated by thin shells covering a wide range of values for $t_r$. We have analyzed the entanglement entropy of an interval
on the unit circle where the dual CFT$_2$ lives. The length of the interval is $\theta\!\in[0,\pi]$, with $\theta\!=\!\pi$ corresponding to the semicircle. The holographic dictionary reduces this problem to evaluating the length of certain bulk geodesics. The results are plotted in Fig.\ref{fig:EE3}a. We observe the same pattern as in Fig.\ref{fig:EE4}a, but now sustained over long cycles. Processes with masses close or above the black hole threshold \eqref{thres} lead to very large $t_r$ and still analogous results, see Fig.\ref{fig:EE3}b. This has the important consequence that even the entanglement entropy of the semicircle remains a long time at its maximum. Contrary to AdS$_4$ processes, two independent time scales emerge. The revival time $t_r$, which grows with the energy density created by the quench until the threshold for fast thermalization is reached. The second one is the time at which the entanglement entropy of half the circle reaches its maximum, which turns out to be very approximately independent of the initial conditions
\be
t_c\approx {\pi \over 2}  \, .
\label{tcollapse}
\ee
This is the time after which some coarse grained observables, such as the entanglement entropy, achieve values characteristic of an ergodic state. In this sense, the subscript in the previous definition stands for collapse.

\begin{figure}[h]
\begin{center}
\includegraphics[width=16cm]{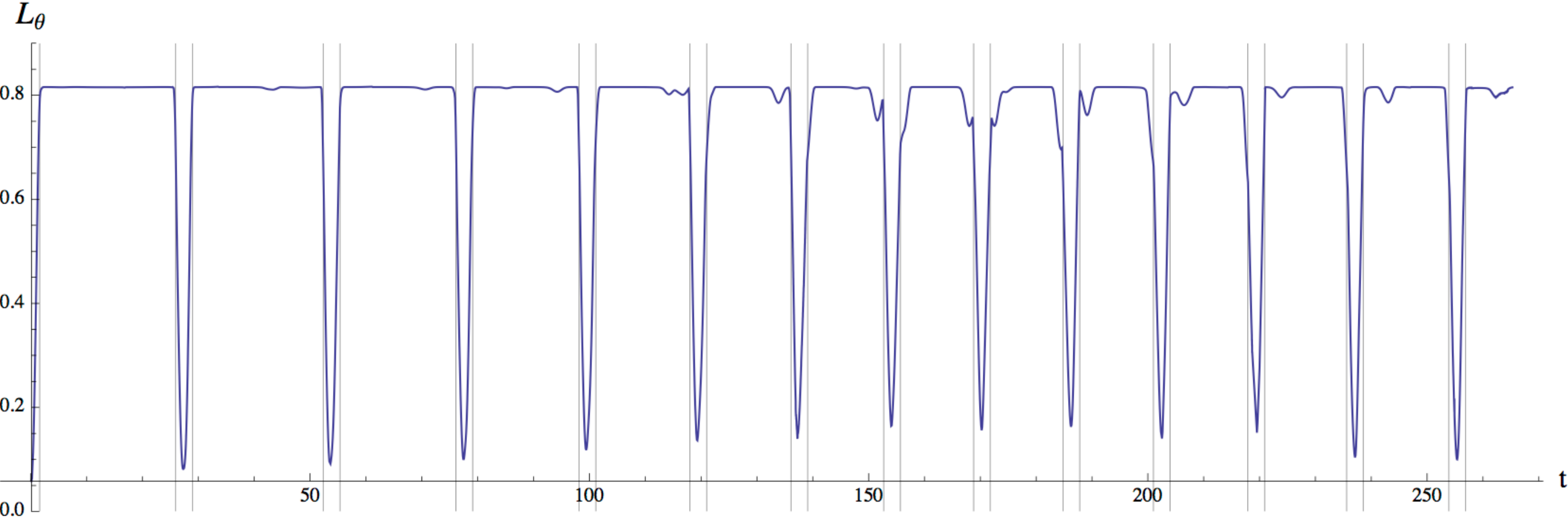}
\end{center}
\caption{\label{fig:EE3long} Entanglement entropy evolution for $\theta\!=\!3$, in a AdS$_3$ process with $\sigma\!=\!0.25$ and ${\cal M}\!=\!1$. The vertical lines mark equal intervals whose extent is  $\Delta t\!=\!\theta$, except for the first line that is located at $t\!=\!\theta/2$.}
\end{figure}

The pattern just described persists along successive revival cycles, in spite that their duration might vary, as shown in Fig.\ref{fig:EE3long}.

\section{Field theory interpretation}

A very simple model was proposed in \cite{Calabrese:2005in} to describe the evolution of 1-dimensional systems after a quantum quench. It assumed that the excitations generated by the quench move as free quasiparticles following classical trajectories according to its group velocity. When the system after the quench is critical, all quasiparticles move at the speed of light, giving rise to a sharp light-cone effect. The early evolution of the entanglement entropy in the finite size CFT$_2$ and CFT$_3$ holographic models we are considering  effectively follows this simple model. Let us assume that entangled excitations are emitted in pairs that move at the speed of light in opposite directions. At time $t=\theta/2$ excitations inside a region of size $\theta$ could only be entangled with those outside, leading to a maximal value for its entanglement entropy. This behavior is reproduced holographically by the implosion of the matter shell \cite{Abajo-Arrastia:2014fma}, as the dotted lines in Fig.\ref{fig:EE4}a and Fig.\ref{fig:EE3}a show.

It is important to stress that the light-cone propagation of entangled excitations was proposed as an effective picture for the evolution of CFT's on an infinite line \cite{Calabrese:2005in}. Indeed a relevant test on the associated holographic models has been showing that it is correctly reproduced \cite{AbajoArrastia:2010yt}. The same simple picture on finite size setups should a priori only apply to the early time evolution, where the size of the system is not relevant. As we have seen,  this is again fulfilled by holographic models. On the contrary, its use to explain finite size effects has to be analyzed in a case by case basis. 

The evolution after a quench of a rational CFT on a circle was studied in \cite{Cardy2014}. Partial revivals of the initial state were found at integer multiples of 
\be
t_r=\pi \, ,
\label{revCC}
\ee
for a circle of unit radius.
This coincides with the time that takes free quasiparticles emitted together in opposite directions and moving at the speed of light to rejoin again, in agreement with the simple propagation picture. However AdS$_3$ processes exhibiting revivals only satisfy  \eqref{revCC} for scalar profiles of very low mass.
Notice that this  does not need to contradict \cite{Cardy2014}. The holographic dictionary maps the classical gravity description with the limit of infinite central charge on the field theory side. Hence the dual CFT will not be a rational one \footnote{We find reasonable to assume that at large but finite central charge the previous conclusion also applies, since otherwise a priori small quantum effects should radically change the classical picture of collapse.}. A consistent picture for the departure of the holographic revival times from \eqref{revCC} calls for an interpretation as an effect of interactions in the quantum field theory. We will give support for it in the following paragraphs.

On generic interacting systems, it is natural to expect that as the energy density created by the quench increases, a fast evolution towards equilibration sets in frustrating the possibility of revivals. The holographic representation of a fast approach to ergodic behavior is the formation of a black hole trapping the complete shell by direct collapse. 
The AdS$_4$/CFT$_3$ models clearly follow this expectation. The number of bounces necessary for gravitational collapse decreases with increasing mass of the scalar profile. Moreover thin AdS$_4$ shells only require bounces before collapse for quite small values of ${\cal M}$ \eqref{mass}, a quantity holographically related to the field theory energy density per species, and for them $t_r\!\approx\!\pi$. This implies that only field theory processes reasonably described by the simple propagation model exhibit revivals. Soon after the effect of interactions starts to play a role, a fast approach to ergodicity sets in.

This is not the case in AdS$_3$/CFT$_2$ models, a fact which from the dual point of view should be related to the strong symmetry properties of 2-dimensional CFT's. Considerably larger ratios of energy density per species to system size are compatible with revivals for them. As result, values of $t_r$ much longer than $\pi$ can be obtained. The evolution of entanglement entropy in Fig.\ref{fig:EE3} leads to the following interpretation. At $t_c\!\approx\!\pi/2$ after the quench the isolated system appears, at the macroscopic level, to have dephased and thermalized. The microscopic dynamics leads to rephasing at a later time
\be
t \approx t_r-{\pi \over 2} \, .
\label{trephase}
\ee
The initial state undergoes a revival at $t_r$, at least in the sense that entanglement returns to be peaked on neighboring degrees of freedom. The evolution before dephasing \eqref{tcollapse} and after rephasing \eqref{trephase} appears to be well described by the free propagation model of \cite{Calabrese:2005in}. 
The fact that $t_r$ increases with the energy density of the initial state, supports linking its value with interaction effects. 

Regarding the holographic dictionary, we obtain the following consistent pattern. The free propagation model provides a good account of the evolution at times corresponding to the implosion and expansion of the matter shell. The interaction effects map to the strong gravitational dynamics generated when the shell reaches minimum size, and scatters against itself before starting to expand again. It is then that the profile of the shell changes, tending to radially focus a fraction of the pulse. This facilitates the formation of a horizon at a subsequent implosion, representing irreversible dephasing on the dual field theory. The large values of $t_r$ for AdS$_3$ shells close or above threshold are explained by the difficulty that these shells find to climb their own gravitational potential, which retains them for a long time at the point of maximal implosion. Hence, we are relating the dynamics which determines the formation or not of a horizon with the dynamics of dephasing-rephasing which leads in the field theory to revivals or to equilibration.

\section{Collapse time}

Remarkably holography allows to model a system which, depending on the initial conditions, exhibits revivals of a quite different nature. For small energy densities they are well described by the free streaming of entangled excitations which rejoin again on a finite space. For larger energies the evolution turns out to bear a stronger resemblance with a series of collapses and revivals of the system wavefunction. We will compare the phenomenology we have found in the latter case with that of well known quantum systems which undergo collapse and revivals in their evolution. Namely, a Bose-Einstein condensate of atoms in an optical trap and a two-level atom coupled to quantized radiation in a cavity.

The behavior of a condensate of atoms with repulsive interactions trapped in a 3-dimensional confining potential has been studied both theoretical \cite{walls} and experimentally \cite{Greiner2002}. The repulsive interactions in the setup of  \cite{Greiner2002} were reasonably described by the simple hamiltonian $H\!=\! {1\over 2} U {\hat n} ({\hat n}-1)$, where ${\hat n}$ is the operator counting the number of atoms. When the system is prepared in a coherent superposition of states with different number of atoms, it undergoes perfect revivals at integer multiples of $t_r\!=\!2\pi/U$. Indeed
\be
|\alpha(t)\rangle=e^{-{|\alpha|^2 \over 2}} \sum_n {\alpha^n \over \sqrt{n!}} e^{-{1 \over 2} Un(n-1) t} |n\rangle \, ,
\label{BE}
\ee
with $\alpha$ a complex number characterizing the coherent state.

\begin{figure}[h]
\begin{center}
\includegraphics[width=7.3cm]{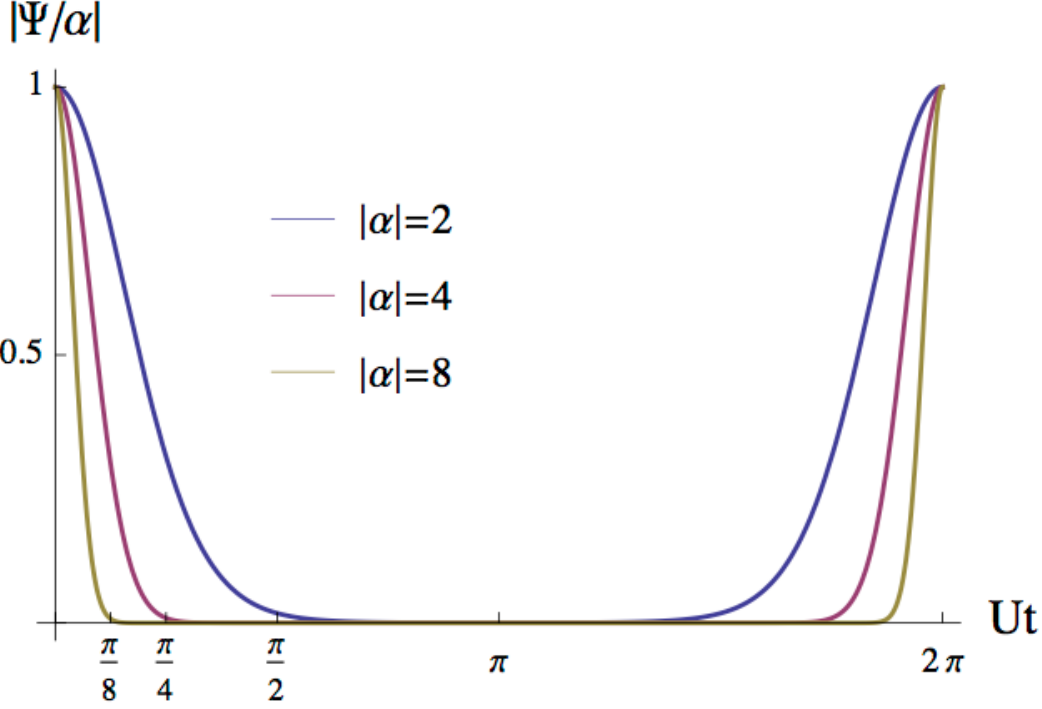}~
\includegraphics[width=7.5cm]{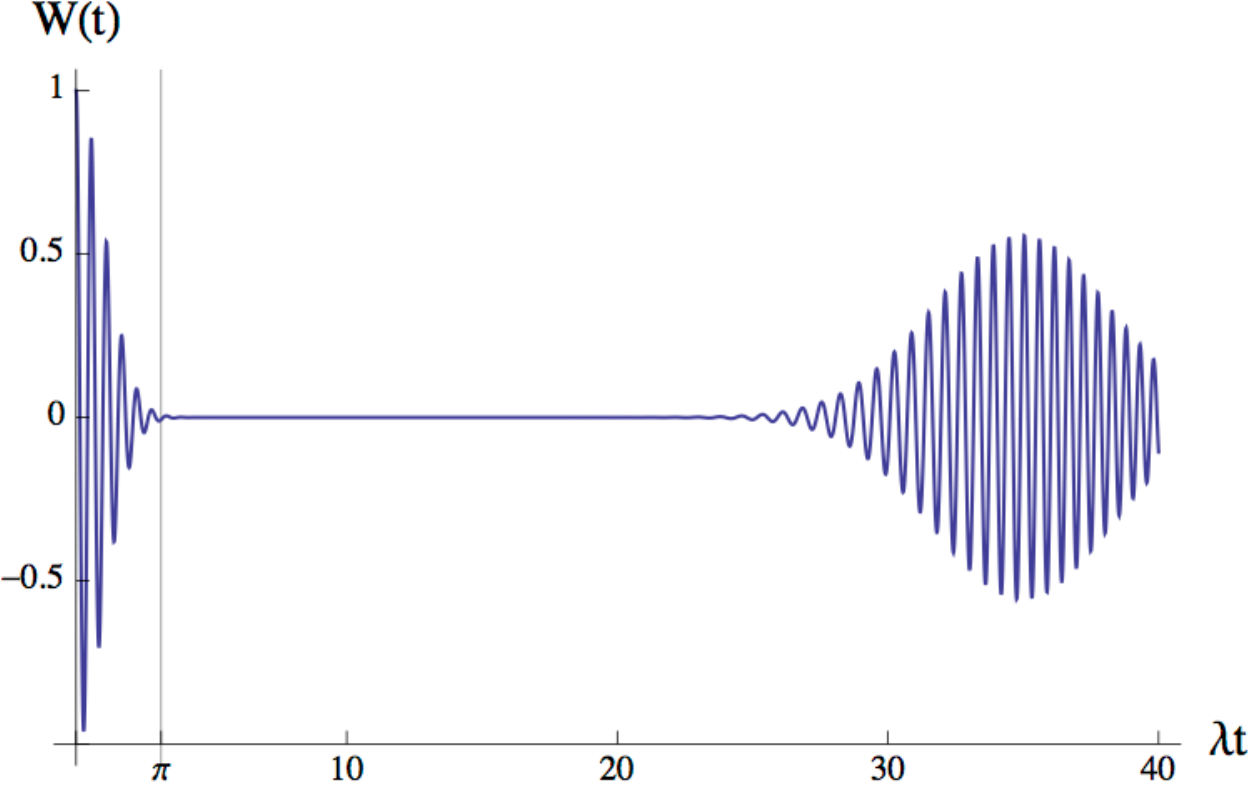}
\end{center}
\caption{\label{fig:rev} Left: Evolution of the matter wave field along a revival cycle for several values of $\alpha$. Right: Evolution of $W(t)$ for $\alpha\!=\!30$. The vertical line signals the collapse time $\lambda t_c\!=\!\pi$.
}
\end{figure}

An important observable is the matter wave field of the condensate, defined by $\Psi(t)\!\equiv\!\langle \alpha(t) | a | \alpha(t) \rangle$. Its mean value represents the fraction of individual atoms that are coherent over the total number of atoms in the trap. Collapse (and revival) of this wave function signals decoherence, again in a coarse grained sense, (and recoherence) of these atoms. The evolution of this observable is straightforward to obtain in the simple case above
\be
\Psi(t)=\alpha \, e^{-|\alpha|^2 (1-\cos U t)} \, e^{i |\alpha|^2 \sin Ut} \, .
\ee
The matter wave function becomes exponentially suppressed at $t_c\!\approx\!\pi/U|\alpha|$ due to the different phases in \eqref{BE}, see Fig.\ref{fig:rev}a. The modulus of $\alpha$ determines the average number of atoms, ${\bar n}\!=\!|\alpha|^2$. Hence the quotient between the revival and collapse times grows with the atom density on the trap
\be
{t_r \over t_c} = 2 \sqrt{\bar n} \, .
\label{quot}
\ee
An qualitatively analogous relation holds in the holographic models. 

Let us review now the behavior of a two level atom in a cavity coupled to quantized radiation. This system is described by the Jaynes-Cummings model \cite{Jaynes:1963zz}. Its hamiltonian is $H\!=\!\omega(\sigma_3/2 + a^\dag a) + \lambda (\sigma_+ a + a^\dag \sigma_- )$, where $a^\dag, a$ are the photon creation and destruction operators and $\sigma_{3,\pm}$  the Pauli matrices referring to the two level atom. The previous hamiltonian refers to the resonant case, where the photon frequency coincides with the energy splitting among the atomic levels. When the radiation field stars in a coherent superposition of states of different number of photons and the atom in the excited state $|+\rangle$, we have
\be
|\Psi(t)\rangle=e^{-{|\alpha|^2 \over 2}} \sum_n {\alpha^n \over \sqrt{n!}} \, \Big[ \cos(\lambda \sqrt{n+1} \, t) \, |+\rangle \, |n\rangle - i  \sin(\lambda \sqrt{n+1}\, t) \, |-\rangle \,
|n+1 \rangle \Big] \, .
\ee

The probability to find the atom in the excited state minus that of finding it in the ground level,  $W(t)\!=\!|\langle+|\Psi(t) \rangle |^2-|\langle-|\Psi(t) \rangle |^2$, 
shows a series of collapse and revivals along its evolution. It starts being one but becomes exponentially suppressed at $t_c \!\approx \!{\pi \over \lambda}$, as can be observed in Fig.\ref{fig:rev}b, loosing the imprint of the initial state. 
At $t_r \approx {2\pi |\alpha| \over \lambda}$ the function $W(t)$ returns to have finite values, which partially reconstructs the initial dominance of excited state. Since the average number of photons is again ${\bar n}\!=\!|\alpha|^2$, collapse and revival times satisfy also in this case \eqref{quot}. In addition now $t_c$ is determined by the properties of the system while $t_r$ grows with the energy, in closer analogy with the holographic models.

\section{Decoding the shell profile}

We will address now an important issue left open above. Namely, the relation of the shell thickness with the time span of the field theory perturbation generating the initial out of equilibrium state. At a qualitative level, we shall provide a map of the shell profile to entanglement properties of the initial state.

\begin{figure}[h]
\begin{center}
\includegraphics[width=7.5cm]{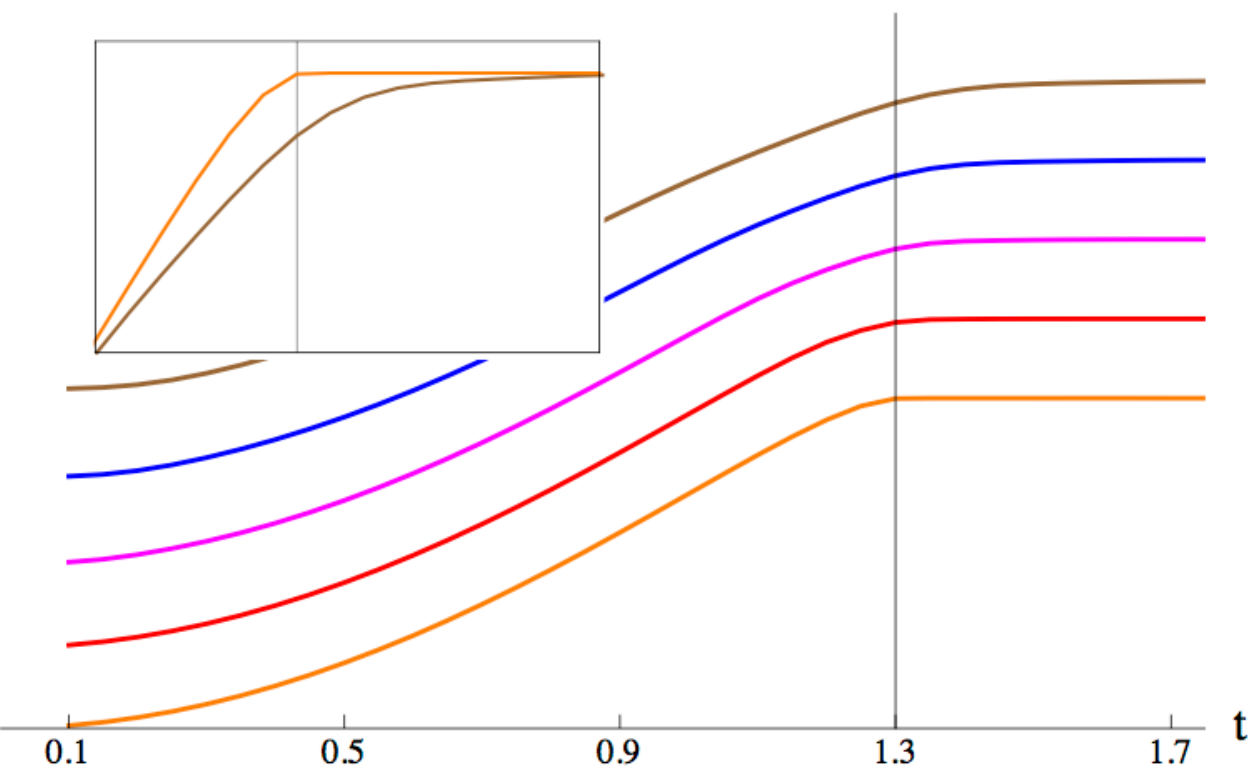}~~
\includegraphics[width=7.5cm]{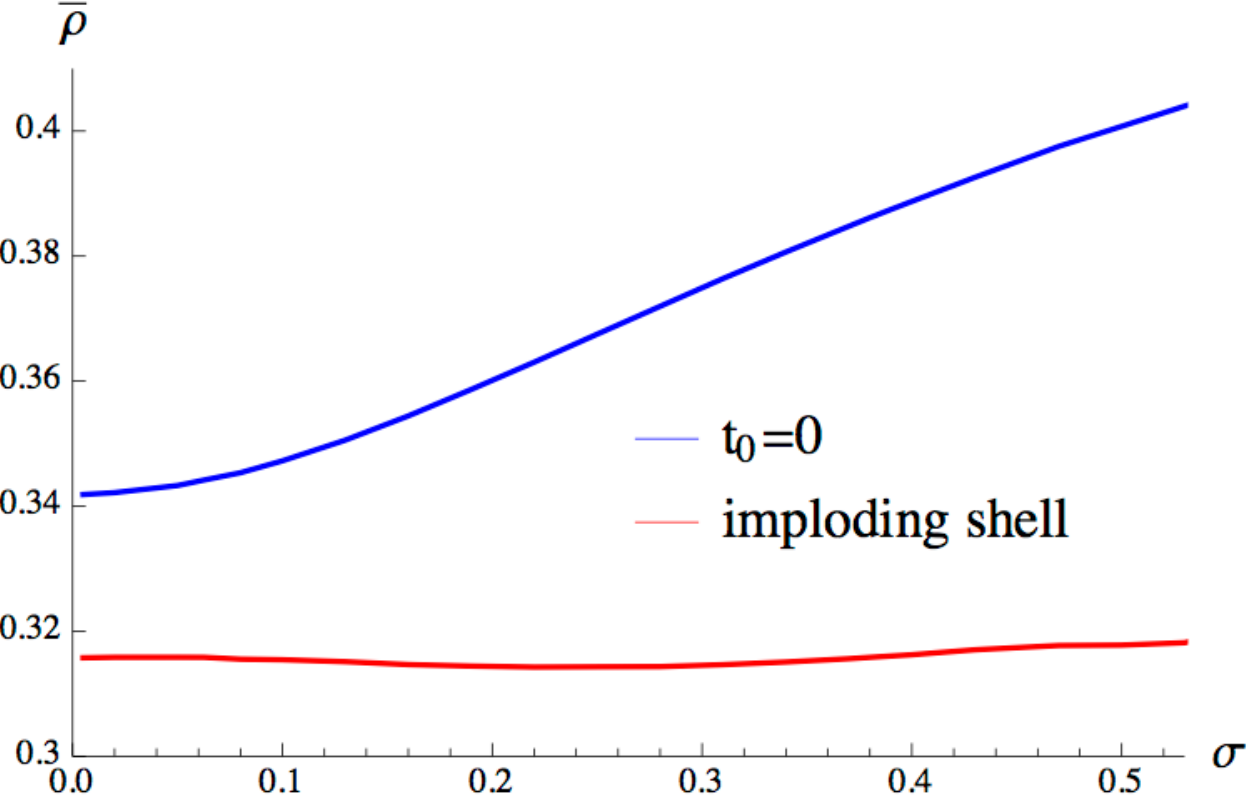}
\end{center}
\caption{\label{fig:amplitude} Left: EE growth for $\theta\!=\!2.6$ along the implosion of AdS$_3$ shells with ${\cal M}\!=\!0.68$ and $\sigma\!=\!0.01,0.05,0.1,0.15,0.2$ from bottom up. We have displaced vertically the lines for the sake of comparison. In the inset, $\sigma\!=\!0.01,0.2$ lines without displacement. Right: Value of $\bar \rho$ for the two choices of $t_0$ described in the text, along the AdS$_4$ transition line to direct collapse, see Fig.\ref{fig:AdS4}a. }
\end{figure}

When the perturbation that brings a system out of equilibrium has a finite time span, it is relevant to take into account that there will be excitations produced at different instants of time. This will cause the entanglement entropy not to reach its maximum until those entangled components emitted last have reached a separation larger that the region considered. If the shell thickness relates to the time span of the perturbation, the mentioned effect should explain the difference between the entanglement entropy plots in 
Fig.\ref{fig:EE4}, holographically derived from AdS$_4$ collapses. Indeed it roughly does. While the EE associated to the thin shell clearly saturates at $t\!\approx\!\theta/2$, that derived from the broad shell does not and instead keeps on growing with a smaller slope to its maximum. A cleaner effect can be observed in Fig.\ref{fig:amplitude}a, where we plot the EE evolution along the first implosion of several AdS$_3$ shells of the same mass and different thickness  $\sigma$. The approach of the EE to its maximum is sharper for the thinner shell, while it smoothness out and requires a longer time for the broader ones.

If the radial thickness of the initial pulse indicates the time span of the perturbation bringing the field theory out of equilibrium, its energy profile should qualitatively measure the density of excitations generated at each instant of time \cite{Abajo-Arrastia:2014fma}. The energy distribution of the shell is given by the integrand in \eqref{mass}, which we will denote by $\rho(t,x)$. Hence, we propose the following qualitative dictionary 
\be
\rho(t_0,x) \to {d \varepsilon \over dt} (t_0+x-\pi/2) \, ,
\ee
where $\varepsilon$ is the field theory energy density at $t\!=\!t_0\!+x\!-\!\pi/2$, and $t_0$ the time at which the perturbation ends. In order to convert radial position into time we used that our metrics are asymptotically AdS, and infalling light rays  in AdS move along $t\!=\!const\!-\!x$. Since stronger correlations should happen between excitations emitted at the same instant of time, it is to be expected that 
\be
{\bar \rho}=\max_x \; \rho(t_0,x) \, ,
\label{rho}
\ee
be a relevant data determining the dephasing dynamics of the system. We will show now that indeed \eqref{rho} plays a more important role than the total energy density 
$\varepsilon(t_0)\!\propto\!{\cal M}$.

The massless scalar field that we are using along the paper to generate a collapse process, because of its second order equations of motion, has two independent modes at the AdS boundary. The leading one is interpreted in the holographic framework as a field theory coupling constant. Hence a perturbation which brings the dual CFT out of the ground state and is easy to model holographically, is to turn on and off this coupling. It correspond to impose boundary conditions at $x\!=\!\pi/2$ where the leading scalar mode is non-zero for a finite interval of time. This creates an imploding matter shell that enters the AdS space from its boundary. In our setup we decided instead to set the leading scalar mode to zero, while providing the shell profile at $t\!=\!0$ as initial data for the collapse. However the initial profile \eqref{profile} has not been generated in the way just described, but rather chosen for its simplicity. As a shortcoming, it is does not give rise to a purely infalling signal, but turns out to source both components that fall and move towards the boundary. For this reason the natural choice $t_0\!=\!0$ is not appropriate. A better criterium is to identify $t_0$ with the instant when all shell components are imploding. 

\begin{figure}[h]
\begin{center}
\includegraphics[width=7.5cm]{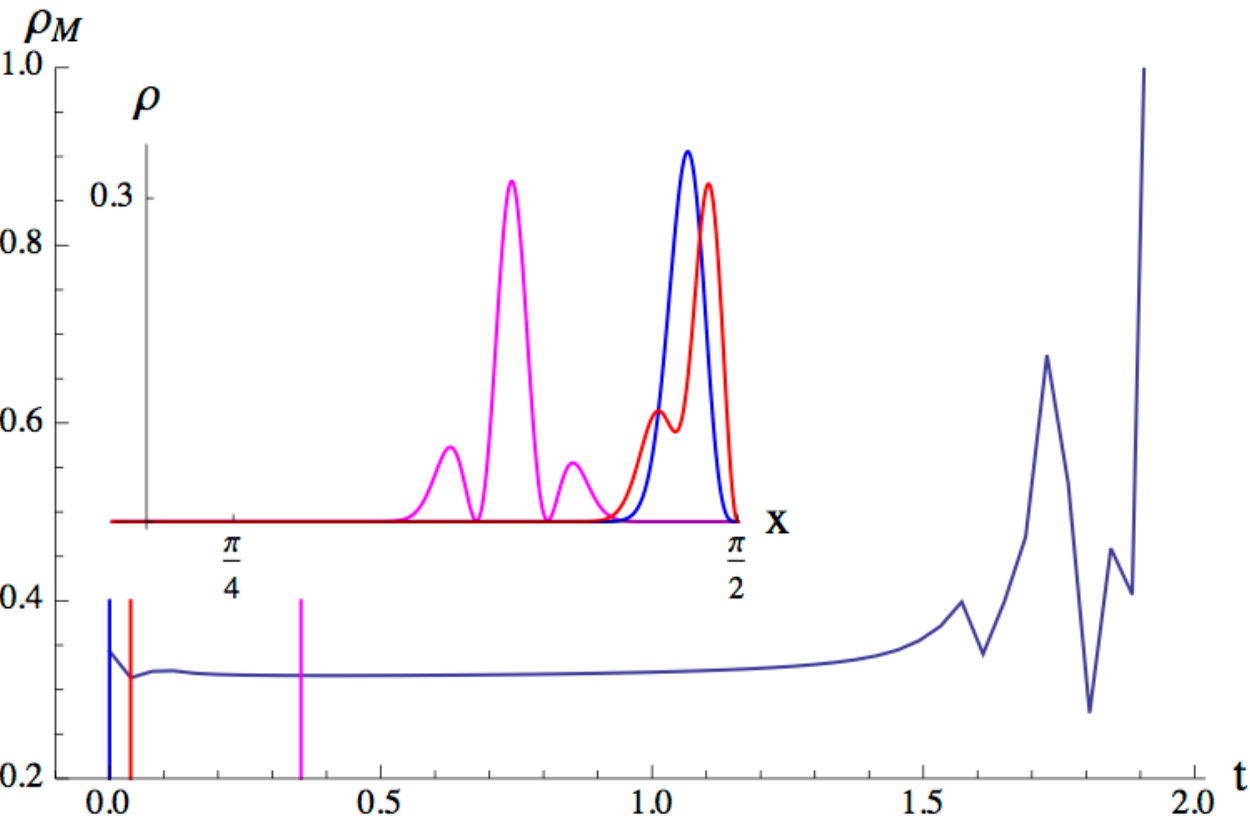}~~
\includegraphics[width=7.5cm]{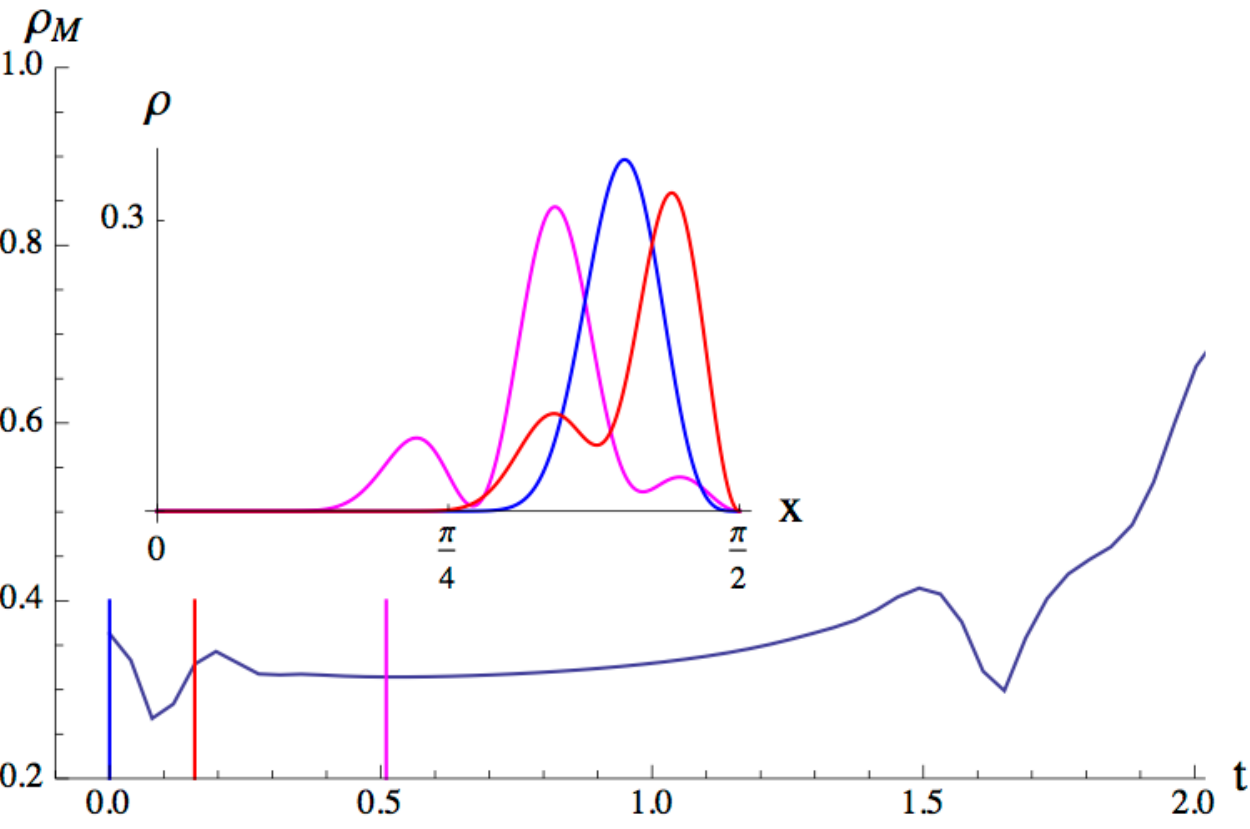}
\end{center}
\caption{\label{fig:rhoM} Evolution of $\rho_M$ along the first implosion of two AdS$_4$ shells on the threshold to direct collapse. Left: $\sigma\!=\!0.05$ and ${\cal M}\!=\!0.023$. Right: $\sigma\!=\!0.22$ and ${\cal M}\!=\!0.09$. On the insets, energy profile at $t\!=\!0$ (blue), at the moment of closest approach to the boundary (red), at time used to evaluate $\bar \rho$ (magenta).}
\end{figure}

We have evaluated $\bar \rho$ along the line separating bouncing geometries from direct collapse in AdS$_4$, see Fig.\ref{fig:AdS4}, for the two choices $t_0\!=\!0$ and $t_0$ determined by the shell infall. While the shell mass goes to zero as the shell thickness vanishes, $\bar \rho$ stays finite in both cases. The variation on the range of $\sigma$ we have studied keeps within 40$\%$ of its value in the thin shell limit for the former choice, while remarkably it stays practically constant for the latter. This is summarized in Fig.\ref{fig:amplitude}b. The red curve has been obtained in the following way. The maximum radial value of the energy density
\be
\rho_M(t)=\max_x \; \rho(t,x) \, ,
\ee
exhibits a plateau starting at the time when the dominant shell components begin to infall, see Fig.\ref{fig:rhoM}a. This plateau lasts longer for thinner shells, but it is well identifiable also in the case of broader ones as shown in Fig.\ref{fig:rhoM}b. We have used the minimum value of $\rho_M$ on that interval as the definition of $\bar \rho$.

Hence according to the above qualitative interpretation of the shell profile, the mass window for bouncing geometries closes down for thin shells because they require a smaller mass for reaching the same maximal value of radial density. Namely, the larger $\bar \rho$ the bigger the initial density of excitations entangled among themselves, and the less stable the system becomes against dephasing. The analysis in AdS$_3$ is less straightforward because the threshold mass \eqref{mass} dominates over any other criterium for collapse. In any case, the fact remains that the mass window above threshold before direct collapse also closes down for thin shells.

\section{Discussion}

The aim of this paper was to propose a holographic description of different paths that a strongly coupled, finite size closed quantum system can undertake towards relaxation. We have analyzed processes of gravitational collapse both in AdS$_4$ and AdS$_3$ as models for quantum quenches in finite size spaces, and  shown that holography accommodates in a  natural way field theory evolutions exhibiting revivals. Depending on the dimension and the energy density created by the quench, the revivals might decay, giving way to a stationary state, or might persist preventing the system to equilibrate. We have observed in our models a clear map between the dynamics of dephasing-rephasing which leads in the field theory to revivals or equilibration, and that of horizon formation on the gravity side.

Concerning the AdS$_4$ collapse processes, there is an important aspect we have not yet touched upon. In order that the formation of black hole can represent the end point of the field theory evolution, the black hole should be stable \cite{Dimitrakopoulos:2014ada}. A stability analysis is especially pressing for black holes formed in collapse processes requiring bounces, whose mass is small compared to the AdS$_4$ boundary size. In an isolated system, the preferred state is that of higher entropy among those with the same energy. Both a black hole and a gas of radiation are possible states of the dual gravity system at fixed mass $M$. The black hole entropy wins over that of a thermal gas whenever \cite{Dias:2011at,Dimitrakopoulos:2014ada}
\be
M>{1 \over G^{4/5}} \, ,
\label{evap}
\ee
where we have assumed the AdS radius to be unity, as done in the rest of the paper. All the imploding matter shells we have considered have 
$M\!\propto 1/G$. Hence in the limit of small Newton's constant, necessary for the validity of the classical gravity description, such masses fulfill \eqref{evap}. This ensures the stability of the associated black holes.

According to the holographic dictionary $1/G$ sets the scale of the dual field theory central charges. It would be thus very interesting to consider small but finite values of the Newton's constant. However the AdS$_4$ geometries exhibiting bounces are characterized by masses  $G M\!\sim\!10^{-1}\!-\!10^{-2}$, such that soon \eqref{evap} could start to fail. It is an open problem whether a quantitative analysis can be carried out including some quantum effects on the gravity side, such as Hawking radiation, and how the phenomenology of holographic revivals would get modified.

Our main results concern the AdS$_3$/CFT$_2$ models. The existence in AdS$_3$ of a mass threshold for the existence of black holes leads to revivals in processes with higher energy than that allowed in AdS$_4$. This fact is most likely related with the strong symmetry properties of 1-dimensional CFT's. The revivals present different features depending on the initial conditions.
When the energy density created by the quench is small, they are well described in terms of the free streaming of entangled excitations \cite{Calabrese:2005in}. As the energy increases interaction effects change the pattern into a series of collapses and revivals with similar properties to those observed experimentally in a variety of systems  \cite{walls,Greiner2002,Jaynes:1963zz,Robinett2004}. A common ingredient for the examples  reviewed in Section 6 was the presence of matter or radiation in a coherent state. Coherent states are regarded as the most classical form of quantum matter. An interesting question is whether the limit of large central charges implicit in the holographic models can be viewed as a related requirement. 

Purely in the gravity context, it is highly desirable to have a criterium determining from the initial matter distribution relevant aspects of its subsequent collapse evolution.
One such aspect is whether a horizon is going to form by direct infall or not. We have studied the transition line between direct collapse and bouncing geometries in AdS$_{3,4}$ for the initial matter profile \eqref{profile}. Remarkably the maximum of the radial energy distribution at the instant when all shell components are imploding, remained practically constant along this line in the AdS$_4$ case. Although rather unmotivated from the gravitational point of view, this provides a criterium governing collapse of the above mentioned type. We have moreover suggested a holographic explanation for it, by interpreting the maximum of the matter energy distribution in terms of the density of strongly correlated excitations in the dual field theory. This provides one more example of how the holographic dictionary can be fruitful for both sides of the correspondence.

\section*{Acknowledgements}
We want to thank Jose L.F. Barbon, Belen Paredes, and Andrzej Rostworowski  for discussions, and German Sierra also for a careful reading of the draft. 
The work of E.daS. is financed by the spanish grant BES-2013-063972.
E.L. has been supported by the spanish grant FPA2012-32828 and SEV-2012-0249 of the Centro de Excelencia Severo Ochoa Programme. The work of J.M.  is supported in part by the spanish grant  FPA2011-22594,  by Xunta de Galicia (GRC2013-024), by the  Consolider-CPAN (CSD2007-00042), and by FEDER. A.S. is supported by the European Research Council grant HotLHC ERC-2011-StG-279579 and by Xunta de Galicia (Conselleria de Educaci\'on). 

\end{document}